\documentclass{article}

\usepackage{arxiv}

\usepackage[utf8]{inputenc}
\usepackage[T1]{fontenc}
\usepackage{hyperref}
\usepackage{url}
\usepackage{booktabs}
\usepackage{amsfonts}
\usepackage{nicefrac}
\usepackage{microtype}
\usepackage{amsmath}
\usepackage{cleveref}
\usepackage{lipsum}
\usepackage{graphicx}
\usepackage{natbib}
\usepackage{doi}
\usepackage{algorithm}
\usepackage{algpseudocode}
\usepackage[table]{xcolor}
\usepackage{multicol}
\usepackage{soul}

\graphicspath{ {./images/} }

\usepackage{chngcntr}
\counterwithin{figure}{section}

\usepackage{mathtools}

\hypersetup{colorlinks=true, linkcolor=blue, citecolor=blue, linktocpage}

\title{Bayesian hierarchical modelling of noisy gamma processes: model formulation, identifiability, model fitting, and extensions to unit-to-unit variability}

\author{
 Ryan Leadbetter \\
  Centre for Transforming Maintenance through Data Science\\
  Curtin University, Kent St, Bentley, WA, Australia 6913 \\
  \texttt{Ryan.Leadbetter@postgrad.curtin.edu.au} \\
   \And
 Gabriel Gonz\'{a}lez Caceres \\
  Centre for Transforming Maintenance through Data Science\\
  Curtin University, Kent St, Bentley, WA, Australia 6913 \\
  \texttt{G.Gonzalez@postgrad.curtin.edu.au} \\
  \And
 Aloke Phatak \\
  Centre for Transforming Maintenance through Data Science\\
  Curtin University, Kent St, Bentley, WA, Australia 6913 \\
  \texttt{Aloke.Phatak@curtin.edu.au} \\
}

\usepackage{setspace}

\begin{document}
\maketitle
\begin{abstract}

The gamma process is a natural model for monotonic degradation processes. In practice, it is desirable to extend the single gamma process to incorporate measurement error and to construct models for the degradation of several nominally identical units. In this paper, we show how these extensions are easily facilitated through the Bayesian hierarchical modelling framework. Following the precepts of the Bayesian statistical workflow, we show the principled construction of a noisy gamma process model. We also reparameterise the gamma process to simplify the specification of priors and make it obvious how the single gamma process model can be extended to include unit-to-unit variability or covariates. We first fit the noisy gamma process model to a single simulated degradation trace. In doing so, we find an identifiability problem between the volatility of the gamma process and the measurement error when there are only a few noisy degradation observations. However, this lack of identifiability can be resolved by including extra information in the analysis through a stronger prior or extra data that informs one of the non-identifiable parameters, or by borrowing information from multiple units. We then explore extensions of the model to account for unit-to-unit variability and demonstrate them using a crack-propagation data set with added measurement error. Lastly, we perform model selection in a fully Bayesian framework by using cross-validation to approximate the expected log probability density of new observation. We also show how failure time distributions with uncertainty intervals can be calculated for new units or units that are currently under test but are yet to fail.

\end{abstract}

\newpage

\tableofcontents

\newpage

\section{Introduction}
Gamma stochastic processes have been widely used in the reliability literature for modelling degradation that evolves monotonically \citep{lawless_covariates_2004}. Once the parameters of the stochastic process have been estimated, the model can be used to forecast the degradation and hence to obtain useful quantities such as failure time and remaining useful life distributions. Maximum likelihood and related methods have often been used to estimate the parameters of degradation models based on stochastic processes, but recent work makes use of the Bayesian paradigm for modelling and inference, especially as simulation algorithms and tools have become more sophisticated and widely available. The development and use of flexible probabilistic programming languages such as BUGS \citep{Lunn_2012}, JAGS \citep{Plummer_2003} and Stan \citep{Stan_2022} has meant that more complex models can be postulated and fitted \citep{McElreath_2020}. Nevertheless, increased power and flexibility---the ability to fit and explore complex models relatively quickly---means that analyses should respect the precepts of good statistical practice: in the Bayesian paradigm, this means an emphasis on thinking carefully about conditional independence assumptions that can make complex models more tractable, which prior distributions should be used, prior and posterior predictive checking, assessing model fit, and the use and dissemination of reproducible code \citep{gelman_bayesian_2020}.

In this paper, we illustrate how:
\begin{enumerate}
    \item the modelling of noisy gamma processes can be simplified by expressing them within a Bayesian hierarchical modelling (BHM) framework \citep{Berliner_1996, cressie_statistics_2011} and encoding uncertainty using conditional probability distributions; 
    \item the basic BHM for a single degradation path can encompass extensions such as unit-to-unit variability;
    \item we can now think critically about how to parameterise the model and specify the distributional forms used for priors and hyperpriors because we are no longer constrained to using conjugate priors, thanks to new Markov chain Monte Carlo methods and statistical computing environments;
    \item the practice of model construction, checking, and comparison, and reproducible research developed in other applied Bayesian fields can be used by the reliability community \citep{gelman_bayesian_2020}.
\end{enumerate}

We begin in Section~\ref{sec:GP} with a brief introduction to the gamma process as it is used for degradation modelling, and to current work on modelling noisy gamma processes. In Section~\ref{sec:BHM_noisyGP} we introduce Bayesian hierarchical modelling and show how to formulate a model for a noisy gamma process within the BHM framework. As well, we discuss the merits of reparameterising the gamma distribution in terms of orthogonal parameters that are interpretable in terms of the average degradation rate and volatility of the gamma process; these parameters allows us to more easily think about how to specify prior distributions for the parameters of a gamma process. We  discuss the equally important step of prior predictive checking, which allows us to specify sensible prior distributions. In Section~\ref{sec:noisy_gp}, we fit the BHM of a noisy gamma process to simulated degradation traces and illustrate some of the challenges of separating the measurement error from the inherent volatility of the gamma processes when the degradation trace is relatively short. This section also introduces posterior predictive checking as a means of assessing how well the BHM fits the data and also contains a discussion of some computational issues that may arise. In Section~\ref{sec:unit-to-unit} we combine the framework introduced in Section~\ref{sec:BHM_noisyGP} and the results of Section~\ref{sec:noisy_gp} to construct and then fit noisy gamma process models to data from several nominally identical units where the degradation exhibits unit-to-unit heterogeneity. We show how to formulate models in the BHM framework that allow us to specify how information is shared (`pooled') or not among the units and to make their structure transparent. After fitting different models to the data, we show how Bayesian cross-validation can be used to determine which of the models is to be preferred. The section concludes by showing how to use simulation to calculate the posterior distribution of the failure time distribution from the Bayesian hierarchical models. We end in Section~\ref{sec:disc_concl} by summarising the main results and pointing the way to future work. The supplementary material for this paper can be found on a the GitHub repository  \citep{code_repo}.

\section{Gamma processes in degradation modelling} \label{sec:GP}

The gamma process, a type of stochastic jump process, is widely used in the reliability literature as a model for monotonically increasing degradation processes. It was introduced to the reliability domain by \citet{abdel-hameed_gamma_1975}, and since then it has been used in many applications, including the modelling of the corrosion of steel coatings, wear of brake pads, erosion of breakwaters, thinning of pressure vessels, and degradation of LED lights \citep{van_noortwijk_survey_2009}. General path models based on linear or nonlinear models are also widely used for modelling asset degradation \citep{ye_2015}, and the Bayesian hierarchical modelling that we describe here has been used recently to fit a hierarchical general path model to pipe thickness measurements along and around the circumference of pipe spools \citep{Tian2019}. 

Consider a sequence\footnote{Strictly speaking, we should distinguish between the symbols used for a random variable and a possible value that it may take. In the development that follows, however, we do not do so for notational convenience, but we believe that doing so will not cause any confusion. Where it is essential to do so, as in Section~\ref{subsec:ftdist}, we make the distinction.} $\{z_i\}$ of noise-free measurements of the degradation of a unit observed at times $t_i$, $i = 0, 1, 2 \ldots, I$. Without loss of generality, we assume that $z_0 = 0$ at $t_0 = 0$. A gamma process \citep{lawless_covariates_2004} models the jumps in degradation between measurements, $\Delta z_i = z_i - z_{i-1}$, as independent samples from a gamma distribution. Thus, we can write that
\begin{equation} \label{eq:GP_general}
    \Delta z_i|\eta(\cdot), \xi \sim \mbox{Ga} \left\{ \eta(t_i) - \eta(t_{i-1}), \xi \right\},
\end{equation}
with rate $\xi$ and shape $\eta(t_i) - \eta(t_{i-1})$, where $\eta(\cdot)$ is a given monotone increasing shape function. The simplest gamma process for modelling degradation is a stationary gamma process, which has a linear shape function  \citep{frenk:2007}, for example, $\eta(t_i) = \beta t_i$. Of course, nonlinear shape functions can be used; however, even when the degradation trace appears to be nonlinear, a time transformation can often be applied so that a stationary gamma process can be fitted. Therefore, in what follows we consider only the stationary gamma process. When using a linear shape function, we can write eq.~\eqref{eq:GP_general} more simply as
\begin{equation} \label{eq:GP_stationary}
    \Delta z_i| \beta, \xi \sim \mbox{Ga} \left( \beta \Delta t_i, \xi \right),
\end{equation}
where $\Delta t_i = t_i - t_{i-1}$.

The gamma process described in eqs.~(\ref{eq:GP_general}) and (\ref{eq:GP_stationary}) can be extended to describe situations commonly encountered in practice, namely, the need to account for measurement error and/or unit-to-unit variability when the degradation of several identical or similar units is being measured. We discuss measurement error below and defer discussing unit-to-unit variability until Section~\ref{sec:unit-to-unit}.

\paragraph{Measurement error (noise)}

Most degradation data collected in industrial settings is contaminated by noise, or error. This noise can be attributed to different sources, including measurement error, instrument noise, placement of sensors, and other environmental factors \citep{ye_2015}. Consequently, models for gamma processes must be extended to account for such noise. 

In an early paper, \citet{kallen_optimal_2005} fit a single parameter gamma process to noisy data by using the additive model $y_i = x_i + \epsilon_i$, where $y_i$ represents the noisy observations, $x_i$ represents the underlying gamma process, and $\epsilon_i$ is independent and identically distributed Gaussian noise. The gamma process is parameterised in terms of the mean wear rate ($\beta / \xi $). They then use the differences of the measured (noisy) jumps, $\Delta y_i = y_i - y_{i-1}$, to formulate the likelihood; consequently, the likelihood is determined by a convolution because the random variable $\Delta Y_i = \Delta X_i + \Delta E_i$ is the sum of the two random variables $\Delta X_i = X_i - X_{i-1}$ and $\Delta E_i = E_i - E_{i-1}$. In addition, calculating the difference of the errors leads to a dependence structure between the $\Delta \epsilon_i$s . To carry out inference, \citet{kallen_optimal_2005} use simulation to approximate the likelihood. \citet{lu_efficient_2013} extended their work by developing a faster method for approximating the likelihood using the Genz transform and a quasi-Monte Carlo method. Their method also allows both of the parameters of the gamma process, $\beta$ and $\xi$, in \eqref{eq:GP_stationary} to be estimated. \citet{peng_2018} use similar methodology.

Building on the work of \citet{kallen_optimal_2005} and \citet{lu_efficient_2013}, \citet{pulcini_perturbed_2016} proposed a way to include degradation-dependent measurement error. Other researchers focused on improving computational efficiency by alternative methods such as deconvolution \citep{rodriguez-picon_reliability_2021} or by using  faster algorithms to approximate the likelihood, for example, approximate Bayesian computing \citep{hazra_approximate_2020, hazra_likelihood-free_2022}. Common to all of these works, however, is a convolution-based likelihood based on a \emph{marginal} model that requires the evaluation of, or approximations to, a complicated multidimensional integral. By contrast, hierarchical modelling based on \emph{conditional} models provides a more straightforward, tractable, and flexible alternative when it is combined with an efficient inferential method. We describe hierarchical modelling in a Bayesian framework in the next section, but first note in passing that \citet{giorgio_perturbed_2019} and \citet{esposito_new_2022} also formulate a conditional likelihood to model a complex noisy gamma process and use maximum likelihood estimation combined with an EM algorithm and particle filtering for estimation and inference.

\section{Bayesian hierarchical modelling of gamma processes with noise}\label{sec:BHM_noisyGP}

Our objective in this section is to outline Bayesian hierarchical (or multilevel) modelling and demonstrate that it simplifies modelling noisy gamma processes. Although we focus on Bayesian inference, it is worth noting that inference for hierarchical models need not be Bayesian; for example, generalized linear models can be cast as hierarchical models containing a random and systematic component \citet[Section~5.6]{dunn2018}, with parameters estimated by maximum likelihood. Nevertheless, Bayesian inference for hierarchical models has several advantages that have been outlined by \citet{reich2019}, including handling missing data, censored data, multiple data sources, and complex hierarchical structures.

In Bayesian inference, we make probability statements about a parameter $\theta$ conditional on observed data $y$ and (implicitly) on values of covariates \citep{gelman2013bayesian}.  To do so requires a model for the joint distribution of $y$ and $\theta$, $P(y, \theta)$, which we can write as $P(y, \theta) = P(y|\theta) P(\theta)$, the product of the sampling distribution $P(y | \theta)$ and the prior distribution $P(\theta)$. By symmetry, the product can also be written as $P(\theta|y) P(y)$, and equating the two products and rearranging leads to Bayes' rule
\begin{equation} \label{BRule}
    P(\theta|y) = \frac{P(y|\theta)P(\theta)}{P(y)},
\end{equation}
where $P(y) = \int P(y|\theta) P(\theta) d\theta$ for continuous $y$.  In eq.~(\ref{BRule}), the prior distribution $P(\theta)$ encodes the analyst's belief about the parameters before s/he has seen the data. After the data have been observed, this prior belief is then updated using Bayes' rule to yield the posterior distribution $P(\theta|y)$, which provides a complete summary of the distribution of the parameter $\theta$ after the data have been observed.

For a given data set, the denominator $P(y)$ can be considered fixed and the unnormalised posterior can be written as $P(\theta|y) \propto P(y|\theta)P(\theta)$. Unfortunately, the posterior is only available in closed form for relatively simple models. Exploring the posterior distribution arising from more complex models requires Markov chain Monte Carlo (MCMC) methods \citep{gilks1996markov}, which have been encoded in probabilistic programming languages such BUGS\citep{Lunn_2012} and Stan\citep{Stan_2022}. These statistical computing environments provide a consistent syntax for expressing hierarchical models, computational engines for implementing MCMC methods, and diagnostic plots to visualize the efficiency of the Monte Carlo methods. 

\subsection{Hierarchical statistical modelling}{\label{Main_hierarhical}}

Hierarchical statistical modelling, as described by \citet{cressie_statistics_2011}, is a formalism for describing and modelling data that is generated from an underlying, or `hidden', process, which is of scientific interest to the analyst. Consistent with its philosophy of breaking down a complex problem into manageable steps, the formalism consists of a hierarchy of levels, each of which is described by a conditional probability distribution. The conditional distributions in each level or (sub-)model arise from a factorization of the joint distribution of the data, process, and parameters, and it can be represented conceptually as
\begin{align*} 
p(\mbox{data, process, parameter}) = \: & p(\mbox{data | process, parameter}) \times \\
& p(\mbox{process | parameter}) \times \\
& p(\mbox{parameter}). \\
\end{align*} 
The first level is the data model, $p(\mbox{data | process, parameter})$, which describes the observation process. The second level in the hierarchy is the process model, $p(\mbox{process | parameter})$. It describes the underlying process that is of scientific interest. The third level in the hierarchy, $p(\mbox{parameter})$, is the parameter model, and in a Bayesian setting refers to the prior distribution. This level encodes how we expect the parameters of the model to behave. Each of these different levels in the hierarchy can also be made up of smaller constituent conditional models. For example, the parameter model may be specified in two steps, as $p(\theta|\phi)$ and $p(\phi)$. As we will see in Section~\ref{sec:unit-to-unit}, doing so allows us to construct a model for the degradation paths of many units that allows us to share, or `pool', information about them.  Setting out the model as we have above makes its structure explicit and transparent. This hierarchical modelling framework is widely used in fields where complex models are common, such as spatio-temporal modelling \citep{wikle_spatio-temporal_2019}, and in many other areas \citep{gelman_bayesian_2020}. In the next section, we show how this framework can be used to model noisy gamma processes.

\subsection{BHM of a noisy gamma process} \label{subsec:BHM_nGP}

Figure~\ref{fig:sim-data} shows a degradation trace generated from an underlying gamma process having a linear degradation rate to which noise has been added. In the discussion below, $y_i$ refers to the measured degradation data at time $t_i$, $i = 0, 1, 2, \ldots, I$. Using the same notation as in Section~\ref{sec:GP}, we denote by $\{ z_i \}$ the values of the underlying gamma process at times $t_i$. As a consequence of the independence of the increments and eq.~\ref{eq:GP_stationary}, $z_i = \sum_{j = 0}^i \Delta z_j$ has a gamma distribution given by $\mbox{Ga}(\beta t_i, \xi)$. 

The first step in constructing the BHM is to postulate a plausible model for the observed data. Here, we assume that \emph{given the value of the gamma process}, the observations are independent of each other; in other words, the $y_i$ are \emph{conditionally} independent. If we know the value of $z_i$, knowing $y_i$ does not provide us with any information about $y_j$, where $i\neq j$. An equivalent way of expressing this is to write $y_i = z_i + \epsilon_i$, where $\epsilon_i$ represents `error' or `uncertainty', and then to assume that the errors are independent of each other. In this work, we will assume that the observations conditional on the underlying process have a Gaussian distribution, and hence the data model is
\begin{align*}
    y_i|z_i, \sigma & \sim \mbox{N}(z_i, \sigma)  && \mbox{data model}
\end{align*}
where $\sigma$ is the standard deviation of the Gaussian distribution. The second layer characterizes the process model by the gamma stochastic process described above and hence can be written as 
\begin{align*}
    z_i & = \sum_{j=0}^i \Delta z_j \\ 
    \Delta z_i | \beta, \xi & \sim \mbox{Ga}(\beta \Delta t_i, \xi) && \mbox{process model}
\end{align*}
The final layer is the parameter model, in which we specify the distribution of the parameters $\beta, \xi$ and $\sigma$. For the moment, we write the distribution in its most general form, as the joint distribution
\begin{align*}
    \beta, \xi, \sigma | \theta & \sim \pi(\theta) && \mbox{parameter model}
\end{align*}
where $\pi(\theta)$ represents the parameters of the joint distribution. Taken together, and with an appropriate inferential mechanism, these three layers make up the Bayesian hierarchical model for a gamma process with noise. In subsequent sections we discuss how to reparameterise the process model so that interpretation of the parameters is more straightforward, how to think about choosing priors, and how choosing these two elements carefully can facilitate more efficient computation. In Section~\ref{sec:unit-to-unit}, we will also discuss how the BHM can be extended to model unit-to-unit variability when similar units are being tested or measured.

\subsection{Reparameterising the gamma process} \label{subsec:param}

The gamma process described above has the density function
\begin{equation}
    f(z_j; \beta t_i, \xi) = \frac{\xi^{\beta t_i}}{\Gamma(\beta)} e^{-\xi x} z^{\beta t_i - 1}, \label{eq:GamDist}
\end{equation}
and the mean and variance, which we denote by $\mu$ and $\sigma^2$, are given by
\begin{equation}
    \mu = \frac{\beta}{\xi}t_i \,\,\,\,\mbox{and}\,\,\,\,\sigma^2 = \frac{\beta}{\xi^2}t_i. \label{eq:GamProp}
\end{equation}

Both the average degradation rate and the variability of the gamma process depend on the parameters $\beta$ and $\xi$. Hence, it is challenging to specify prior distributions of $\beta$ and $\xi$ so as to separate their effects on the stochastic process. From the perspective of the user, it is desirable to reparameterise the gamma process so that the new parameters have clear interpretations and effects. In addition, if they are \emph{orthogonal} \citep{cox_reid_1987}, there are several desirable statistical consequences for estimation, inference, and computation.

One such reparameterisation is in terms of the mean $\mu$ and coefficient of variation $\nu = \sigma/\mu = 1/\sqrt{\beta}$: the mean represents the average degradation rate, whereas the coefficient of variation describes the inverse of the `signal-to-noise' ratio of the stochastic process. For the user, therefore, $\mu$ and $\nu$ have a more intuitive interpretation than the shape and the rate. Furthermore, using a result due to \citet{huzurbazar_1956}, it is straightforward to show that these parameters are also orthogonal. (We note in passing that orthogonal parameterisations are not unique; the mean $\mu$ and shape $\beta$ are also orthogonal \citep{huzurbazar_1956}.)

Substituting $\mu$ and $\nu$ in the expression for the distribution of the increments in the process model in Section~\ref{sec:GP} yields
\begin{align} \label{eq:GP_stationary_reparam}
    \Delta z_i|\mu, \nu & \sim \mbox{Ga} \left( \frac{\Delta t_i}{\nu^2}, \frac{1}{\mu \nu^2} \right).
\end{align}
We use this reparameterisation in the remainder of this paper. \citet{kallen_optimal_2005} also use the shape and coefficient of variation, pointing out that it can be easier for a plant engineer to interpret them. They do not, however, exploit their orthogonality, preferring to fix the value of $\nu$ in their analysis instead of estimating it.

\subsection{Prior distributions and prior predictive checking}
\label{subsec:prior}

The prior distribution in the parameter model summarizes our beliefs about the parameters. There are two different ways in which this information is encoded: the choice of distribution, and the values of the hyperparameters. Before the advent of contemporary sampling algorithms, Bayesian analysis relied on conjugate prior distributions, or convenient prior distributions that facilitated the use of Gibbs samplers or conventional Metropolis-Hastings algorithms \citep{gilks1996markov}. However, with the development of more efficient sampling algorithms such as Hamiltonian Monte Carlo \citep{betancourt_conceptual_2017}, we are no longer limited by such requirements and can select priors that reflect our state of knowledge, facilitate efficient computation, and that can be justified and evaluated in a principled way.  

In the degradation modelling literature, a gamma distribution is often used as the prior distribution for the rate parameter $\xi$ of the gamma process \citep{lawless_covariates_2004} and also for the shape parameter \citep{rodriguez-picon_degradation_2018}. It is well known that a gamma prior on the rate parameter is conditionally conjugate\citep{Pradhan_estimation_2011}, and its use leads to analytically tractable results, as \citet{lawless_covariates_2004} show. Nevertheless, little work has been done to assess whether other prior distributions might be more appropriate. The gamma distribution has a heavy tail, and its use can lead to MCMC chains that converge very slowly or that are highly autocorrelated; moreover, it can lead to physically implausible realizations of the gamma process, as we demonstrate below.

Evaluation of prior distributions is one step in the `Bayesian workflow' \citep[Figure ~1]{gelman_bayesian_2020}, a series of steps that begins with postulating a model and checking its plausibility, and ends with model checking and evaluation, a step that may lead back to the beginning if inference and/or model fit are inadequate. In the Bayesian workflow, the plausibility of a parameter model is assessed by \emph{prior predictive checking}. Because Bayesian models with proper priors for all parameters yield a proper prior marginal distribution of the data, such models are \emph{generative} and we can simulate data from them and then assess whether they are consistent with domain knowledge, for example, the expected magnitude of the generated data \citep{gabry_vis_2019}. In the terminology of \citet{gabry_vis_2019}, priors that when combined with the likelihood lead to simulated data that could be plausibly observed are known as \textit{weakly informative priors}. We illustrate prior predictive checking of a noise-free gamma process using three sets of priors for its parameters: first, `conventional' priors, $\mbox{Ga}(1, 0.001)$ and $\mbox{Ga}(0.001, 0.001)$, that are widely used in the literature for the both shape and rate parameters of the usual parameterization of a noise-free gamma process in eq.~(\ref{eq:GP_stationary}), and second, priors on $\mu$ and $\nu$ in the alternative parameterization of eq.~(\ref{eq:GP_stationary_reparam}). All three sets of priors yield an average degradation rate of 1 unit per unit time. 

In contrast to the weakly informative priors suggested by \citet{gabry_vis_2019}, the distribution $\mbox{Ga}(\epsilon, \tilde{\epsilon})$, where $\epsilon, \tilde{\epsilon}\longrightarrow 0$, is often used as a noninformative prior distribution, especially in mixed linear models, where it is a conditionally conjugate prior for the precision \citep[p.~33]{hodges_2014}. In addition, as we pointed out above, the gamma distribution is conditionally conjugate for the rate parameter: if $\{ z_i \}$, $i = 1, 2, \ldots, n$, represents an independent sample from $\mbox{Ga}(\beta, \xi)$, then the conditional distribution of $\xi$ given $\beta$ and the data is $\mbox{Ga}(n\beta + \epsilon, \sum_{i=1}^n z_i + \tilde{\epsilon})$ when the prior distribution of $\xi$ is $\mbox{Ga}(\epsilon, \tilde{\epsilon})$. Hence, when $\epsilon$ and $\tilde{\epsilon}$ are both small, the prior adds very little information, but it is noninformative \textit{with respect to the rate parameter only}; furthermore, inferences about $\xi$ may be sensitive to the values of $\epsilon$ and $\tilde{\epsilon}$ in data sets where small values of $\xi$ may be possible \citep[p.~130]{gelman_bayesian_2020}. When $\mbox{Ga}(\epsilon, \tilde{\epsilon})$ is used for \textit{both} parameters, we can no longer assume that the joint prior will be noninformative and therefore must evaluate it to determine whether it is indeed diffuse. For further discussion on the consequences of using $\mbox{Ga}(\epsilon, \tilde{\epsilon})$ as a prior distribution and guidance on using more sensible alternatives, see \citet{hodges_2014} and \citet{gelman_bayesian_2020}.

\begin{figure}
    \centering
    \includegraphics[width=0.95\textwidth]{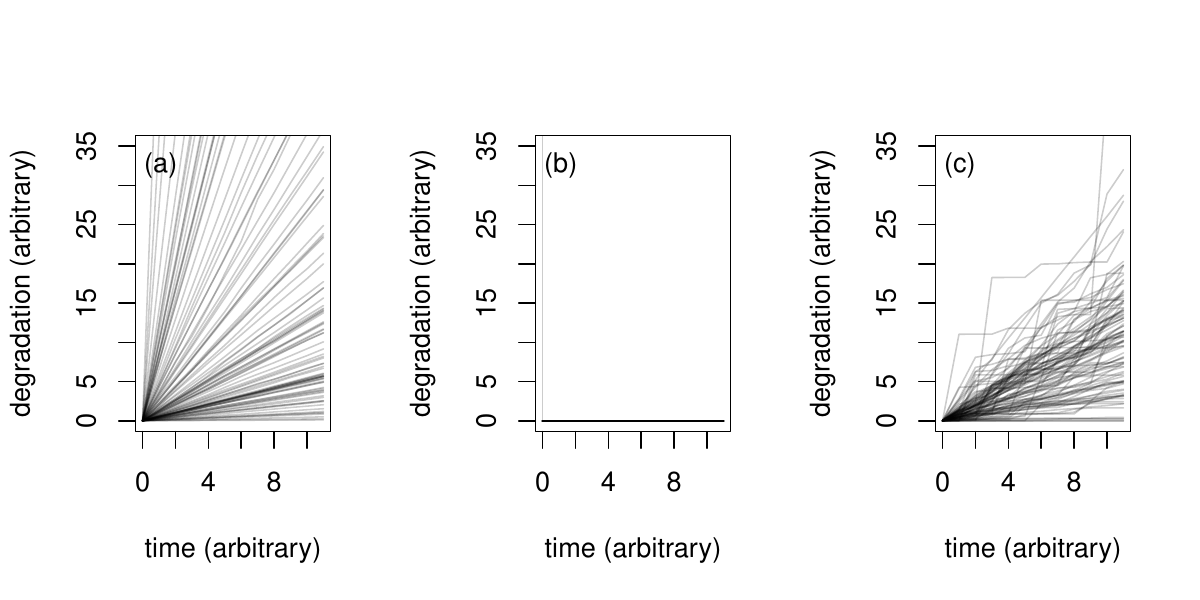}
    \caption{One hundred realizations from prior predictive distributions of a noise-free gamma process with mean degradation rate of 1 unit per unit time and different prior distributions for parameters: (a) shape, rate $\sim \mbox{Ga}(1, 0.001)$, (b) shape, rate $\sim \mbox{Ga}(0.001, 0.001)$, and (c) parameterization mean/coefficient of variation---see text for details.}
  \label{fig:ppc}
\end{figure}

According to \citet{gabry_vis_2019}, we should aim to select priors---weakly informative priors---that will lead to plausible simulated data that may have some mass around extreme, but not completely implausible, realizations. Figure~\ref{fig:ppc}(a) and (b) show 100 draws from the prior predictive distribution of a noise-free gamma process when both the shape and rate parameters are assigned the prior distribution $\mbox{Ga}(1, 0.001)$ (Fig.~\ref{fig:ppc}(a)) or $\mbox{Ga}(0.001, 0.001)$ (Fig.~\ref{fig:ppc}(b)). In Fig.~\ref{fig:ppc}(a), we can clearly see that the degradation traces resulting from a $\mbox{Ga}(1, 0.001)$ prior distribution are all nearly linear, without the jumps expected of gamma processes; furthermore, many of the rates of degradation are unrealistically high and unrealistically low. In Fig.~\ref{fig:ppc}(b), where a $\mbox{Ga}(0.001, 0.001)$ prior is used, most of the prior predictive distribution has mass around implausibly low values of the average rate, and there is one unrealistically steep degradation trace. As we pointed out earlier, the gamma distribution is highly skewed and has heavy tails; consequently, depending on the values of the shape and rate, the prior can place mass on high, low, or both high and low values, resulting in simulated data that simply could not be observed in practice.  By contrast, the reparameterization of the gamma process in terms of the mean $\mu$ and coefficient of variation $\nu$, along with appropriate prior distributions for these parameters, leads to plausible realizations, as Fig.~\ref{fig:ppc}(c) shows. We now turn to the details of this parameterization and its associated prior distributions.

To specify independent prior distributions of the parameters $\mu$ and $\nu$ in the GP model in  \eqref{eq:GP_stationary_reparam}, we adopt the approach introduced by \citet{Simpson_2017}: design priors that favour simpler models over more complex ones and that are consistent with domain knowledge. The mean $\mu$ controls the average degradation rate, similar to the action of the slope parameter in a linear degradation path model. We have no reason to believe that the variability about the mean degradation rate would be asymmetric, so a Gaussian distribution with a small standard deviation is both appropriate and convenient. The coefficient of variation $\nu$ is a measure of the volatility of the degradation process, and although we might expect some heterogeneity in the wear rate as degradation progresses, we do not expect the wear rate to be extremely volatile. Hence, we use a truncated Student $t$-distribution with 3 degrees of freedom as a prior for $\nu$. It places a large mass near zero but still allows the posterior distribution to move away from zero. In addition, it has lighter tails than a gamma distribution, and consequently does not give too much weight to extremely volatile degradation paths. 

Figure~\ref{fig:ppc}(c) shows realizations from the prior predictive distribution using independent priors $\mbox{N}(1, 0.5)$ and a $t$-distribution with 3 degrees of freedom for $\mu$ and $\nu$, respectively. We can see that the degradation traces have the appearance of paths expected from a gamma process, that is, there are discrete jumps between time points, in contrast to Fig.~\ref{fig:ppc}(a), where all the traces are straight lines. Furthermore, more than half the degradation values at the end (eleventh time point) are between 6 and 16, as would be expected when the degradation varies around one. Finally, although there are some extreme realizations, there are only one or two that are completely implausible. 

To fully specify the model, we also need to specify a prior for the standard deviation of the measurement error, $\sigma$. Following the recommendations of \citet[Chapter~17]{gelman_bayesian_2020}, we use a vague $\hbox{Uniform}(0, A)$ prior for $\sigma$, where $A$ is chosen to be large relative to the expected scale of $\sigma$. We use such a vague prior for demonstration purposes. However, in practice, an analyst should have a reasonable grasp of the scale of the measurement error and should be able to specify a weakly informative prior. Because our prior on $\sigma$ is so vague, we do not include the measurement error in the prior predictive checking in Fig.~\ref{fig:ppc}.

Of course, had we used different values of the hyperparameters in the prior distributions of $\mu$ and $\nu$, the appearance of the degradation traces would have been different too. In particular, had we chosen, for example, to use a truncated Gaussian distribution with a large standard deviation as the prior distribution of $\mu$, then physically implausible traces would have undoubtedly been generated. Nevertheless, when using prior predictive checks to evaluate priors and to find sensible ones, the idea is \textit{not} to try different values of the hyperparameters until the realizations are concentrated around the data that we are analysing; instead, as \citet{gabry_vis_2019a} write, the analyst ``should have enough familiarity with the subject matter to look at prior predictive simulations \ldots without needing to make direct comparisons with the data that will be used for model fitting.'' They go on to say that ``a \textit{reasonable} [our emphasis] prior is a prior that yields a reasonable prior data-generating process, not that the researcher should tailor the prior to suit the particular observations in hand.''

\section{Fitting the noisy gamma process to small samples}
\label{sec:noisy_gp}

Before extending the model for a noisy gamma process to incorporate unit-to-unit variability (Section~\ref{sec:unit-to-unit}), we first fit the noisy GP model to a single simulated degradation path. Simulation is an important step in understanding a Bayesian model. If the model and inferential procedure cannot reproduce the true parameter values when the data-generating mechanism is known, there is little hope that it can do so for real data \citep{gelman_bayesian_2020}. The single simulated path example we discuss below shows that a noisy GP is more difficult to fit than a noise-free GP model: when the sample size is small, the parameters describing the variance of the measurement error and the volatility of the underlying gamma process cannot easily be separated because there is not enough information in the data to do so. To demonstrate this problem with identifiability, we simulate and then analyse a BHM of two data sets: one `large' data set consisting of 20 simulated noisy degradation measurements, and another `small' data set that is a subset of 10 points from the large one. We fit the BHM of the noisy GP outlined in Sections~\ref{subsec:BHM_nGP} and \ref{subsec:param} to these two data sets and evaluate their performance. Specifically, we investigate whether the models can reproduce true parameter values and degradation path, and how efficiently they sample from the posterior. A detailed analysis and complete workflow containing all of the R code is available in the Supplementary Material accessible through a GitHub repository \citep{code_repo}.

\subsection{Data simulation}\label{subsec:datasim}
We generate fictitious data from the Bayesian hierarchical model for the noisy gamma process outlined in Section~\ref{subsec:BHM_nGP}, using the priors discussed in Section~\ref{subsec:prior}. It is summarized in Fig.~\ref{fig:BHM-nGP}, where we use the superscript $(+)$ to denote a truncated distribution whose lower bound is zero; furthermore, we use the location-scale form of a $t$ distribution with $df$ degrees of freedom, written as $t_{df}(\hbox{location}, \hbox{scale})$. The `large' dataset was generated by first specifying reasonable values of the parameters: $\mu = 10$, $\nu = 1.119$, and $\sigma = 4$. Then, 20 time increments between measurements were obtained by randomly sampling from a $\mbox{Unif}(0.8, 1.3)$ distribution. Next, we sampled 20 jumps in degradation from $\mbox{Ga}(\Delta t_i/\nu^2, 1/\mu\nu^2)$ and calculated the cumulative sum to obtain the underlying, noise-free degradation path $z_i$, where $z_0 = 0$ at $t_0 (=0)$. Finally, we added Gaussian noise with standard deviation $\sigma = 4$ to the underlying degradation path to get the noisy observations. The fictitious degradation process and noisy observations are shown in Fig.~\ref{fig:sim-data}. The `small' dataset is created from the `large' dataset by randomly selecting ten of the twenty noisy observations; these are the red points in Fig.~\ref{fig:sim-data}. 

\begin{figure}[t]
\begin{align*} 
    y_i|z_i, \sigma & \sim \mbox{N}(z_i, \sigma)  && \mbox{data model} \\
    \Delta z_i|\mu, \nu & \sim \mbox{Ga} \left( \frac{\Delta t_i}{\nu^2}, \frac{1}{\mu \nu^2} \right) && \mbox{process model} \\
    \mu & \sim \mbox{N}^{+}(10, 10) && \mbox{parameter model} \\
    \nu & \sim t_2^{+}(0, 1) \\
    \sigma & \sim \mbox{Unif}(0, 100) 
\end{align*}
\caption{Bayesian hierarchical model for fitting a noisy gamma process to the simulated data in Section~\ref{sec:noisy_gp}.}\label{fig:BHM-nGP}
\end{figure}

\begin{figure}[t]
    \centering
    \includegraphics[width=0.6\columnwidth]{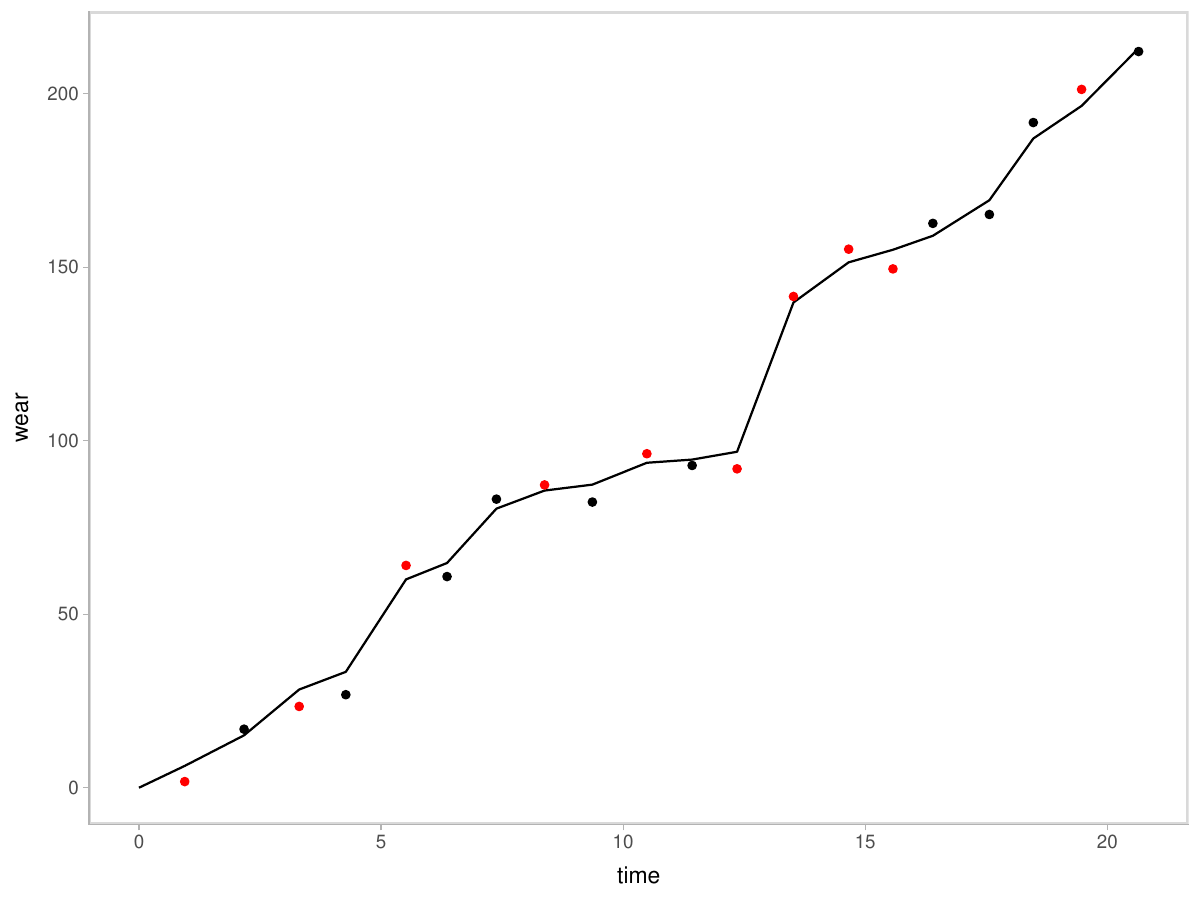}
    \caption{Simulated degradation data. The black line is the underlying degradation path. The black and red points together constitute the `large' dataset of noisy observations. The red points are the subset of observations that make up the `small' data set.}
    \label{fig:sim-data}
\end{figure}

\subsection{Computation}

To fit the noisy GP model to the simulated data and then generate samples from the posterior distribution, we use the No-U-Turn sampler implemented in the probabilistic programming language \textit{Stan} \citep{Stan_2022}. The No-U-Turn sampler is an adaptive variant of Hamiltonian Monte Carlo; a more detailed explanation of how it works can be found in \citet{Hoffman_2014}. To generate the results described below, we generated $28,000$ samples from the posterior distribution using four chains of $10,000$ iterations each, with a burn-in of $3,000$ iterations and no thinning.

\subsection{Results and Diagnostics}

Fitting the model in Fig.~\ref{fig:BHM-nGP} to the two data sets yields multi-dimensional posterior distributions from which the marginal distributions of the parameters of interest can be obtained. Our objective here is to determine how the size of the dataset affects how well the BHM performs, and we therefore assess how well these marginal posterior distributions recover the true values of the parameters. The results show that when the model is fitted to the large dataset, it is able to recover the underlying parameter values; when there are fewer observations, the model fails to do so because it is unable to disentangle the observation noise from the volatility of the gamma process.

\begin{figure}
    \centering
    \includegraphics[width=0.8\columnwidth]{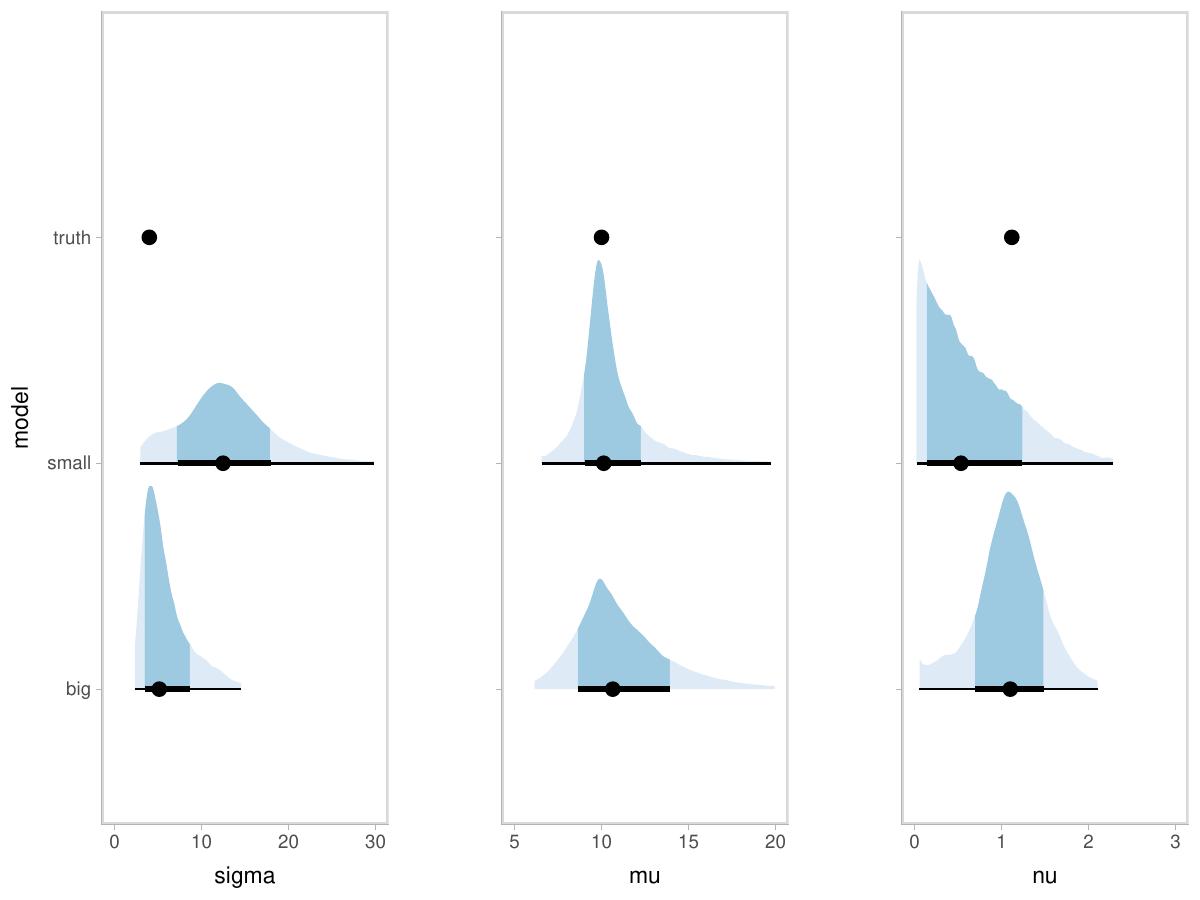}
    \caption{The marginal posterior distributions of the parameters $\sigma$, $\mu$, and $\nu$ when the BHM of Fig.~\ref{fig:BHM-nGP} is fitted to the `large' and `small' simulated data in Fig.~\ref{fig:sim-data}. The points and intervals shown in each distribution represent, respectively, the median and $95\%$ and $66\%$ credible intervals. The values used to simulate the data are shown in the top row.}
    \label{fig:marginal-post}
\end{figure}

Figure~\ref{fig:marginal-post} shows the values of $\mu$, $\sigma$, and $\nu$ used to generate the data in Fig.~\ref{fig:sim-data}, along with the marginal posterior distributions of these parameters conditioned on the small and large datasets. Also shown are the median and 66\% and 95\% credible intervals of each posterior distribution. Clearly, the model fitted to the large dataset has successfully recovered the true values: the posterior medians are almost identical to the true values, and the variability is relatively small. By contrast, the model fitted to the small dataset overestimates the value of $\sigma$ and underestimates $\nu$. In addition, the marginal densities of $\sigma$ and $\nu$ appear to be multi-modal, with a primary mode at $11$ for $\sigma$ and zero for $\nu$, and a secondary mode around the true values of the parameters. Additional simulations shown in Supplementary Material 1 confirm the results shown in Fig.~\ref{fig:marginal-post}---using a larger dataset yields parameter estimates that are closer to the underlying parameter than using a smaller one. 

\begin{figure}
    \centering
    \includegraphics[width=0.8\columnwidth]{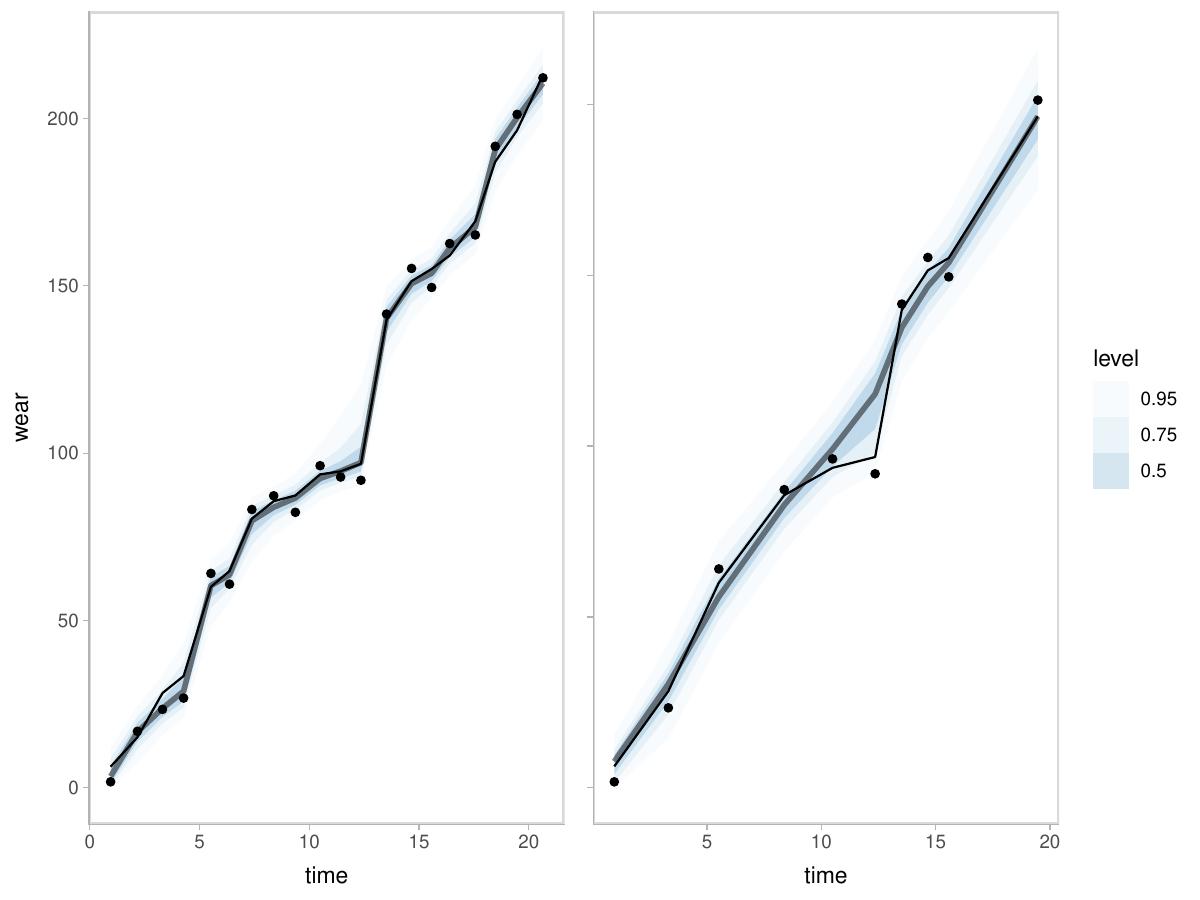}
    \caption{The posterior predictive distribution of the filtered degradation path compared to the true degradation path when the model is fitted using the big (left) and small (right) data set. The median of the predictive distribution is shown as a grey line and the corresponding uncertainty intervals as blue ribbons.}
    \label{fig:ppd-filtered-degradation}
\end{figure}

Although the marginal distributions are useful in identifying whether the BHM can recover the true parameter values, examining the posterior predictive distribution of the filtered degradation path (the $z_i$s) provides us with greater insight about the behaviour of the model fitting. The posterior predictive distribution of the filtered degradation measurements is shown in Figure~\ref{fig:ppd-filtered-degradation}. The thick grey line is the median of the posterior predictive distribution; additional quantiles are shown in different shades of blue-grey. The underlying degradation path and noisy observations are shown in black. The posterior from the large dataset has recovered the true path well: the median path follows the actual path almost exactly, with uncertainty bands that are narrow enough to be useful. However, as was the case with parameter values, the median path calculated from the posterior distribution from the small data set has not recovered the true path. In addition, the uncertainty intervals are much wider. The median path from the small dataset is a nearly straight line through the data points; this `oversmoothing' leads us to conclude that the two modes of the marginal distributions of  $\sigma$ and $\nu$ are associated---when $\nu$ approaches zero and the path from the gamma process becomes nearly linear with only very small jumps, $\sigma$ is overestimated to compensate for the unaccounted-for volatility of the degradation path.

In addition to the poor inference from the small-data posterior of our BHM, sampling from it is inefficient and poorly behaved. In the case of fitting the model to simulated data, where we can be sure that the model is properly specified and implemented, poorly behaved sampling is often a sign of a deeper issue with the model. One advantage of HMC and its variants is the useful within-chain diagnostics that help to not only diagnose when sampling is inefficient but also to investigate why. In particular, divergent transitions \citep{betancourt_conceptual_2017}, or divergences, occur when the leap-frog algorithm used to approximate the Hamiltonian trajectories is not able to explore the tight local curvature in the posterior, and the chain energy races off to infinity. Hence, these divergences tend to be indicate problematic areas in the posterior. \citet{betancourt_conceptual_2017} calls these problematic geometries in the posterior `degeneracies'. The divergences that occurred during the sampling from the small-data posterior are concentrated in deep funnel-shaped degeneracies in the joint distribution of the degradation jumps, $\Delta z_i$, and $\log\nu$, as Fig.~\ref{fig:small-data-pairs} in Appendix~\ref{sec:ApdxA} shows. Funnels occurs around the value $\Delta z_i = \Delta t_i \times \mu$ as $\nu$ shrinks to zero, and they exist between all the $\Delta z_i$s and $\log\nu$. On the other hand, when sampling from the large-data posterior the degenerate behaviour is almost completely eliminated by the extra information in the additional data. The degenerate behaviour in the posterior is more deeply explored in Supplementary Material 1.

\subsection{Discussion on non-identifiability}

This simple example of fitting the noisy GP model to a single simulated noisy degradation path shows that the MCMC algorithm yields degeneracies and that the inference yields biased estimates of $\sigma$ and $\nu$ when the dataset is small. This degenerate behaviour of inference from the small-data posterior is a result of what \citet{betancourt_2020} refers to as `pre-asymptotic non-identifiability'.

Because variation in the noisy degradation signal can be a result of both the randomness of jumps of the GP and the randomness of the measurement error, it can be difficult to separate these two sources when there is only a small number of observations. In the small dataset, the data do not strongly inform the parameters $\sigma$ and $\nu$, and it is therefore difficult to distinguish between competing models---the noisy gamma process, and one where $\nu$ approaches zero---as reflected in their multi-modal posterior distributions.  Using the terminology of \citet{betancourt_2020}, we can say that these parameters are pre-asymptotically non-identifiable. In Supplementary Material 1, we show how adding information to the analysis, either through a stronger prior or additional information that specifically informs one of the pre-asymptotically non-identifiable parameters, eliminates problems with inference or sampling for the small data set. Adding information into the analysis that helps to inform one of the pre-asymptotically non-identifiable parameters makes it possible to distinguish between the competing models and therefore makes fitting the noisy GP better behaved. Another way of adding additional information so that inference and sampling is less problematic is by pooling information between similar units, the topic of Section~\ref{sec:unit-to-unit}.

In the context of real noisy degradation data, there is no way of checking if there is enough information in the data to properly identify the model. Therefore, practitioners applying the noisy GP model should use as much information as they have available to them. This includes encoding their domain-expert knowledge into the prior distributions of the parameters, rather than choosing a default non-informative prior; incorporating supplementary data that informs only the measurement error; and modelling the degradation of groups of similar units jointly as to `borrow' information. In the next section, we model the degradation paths from multiple units while assuming that the measurement error is the same for all units. Doing so drastically improves the problems with identifiability and MCMC sampling that we have seen in this section. However, in order to do this, we must choose a suitable joint model for the units.

\section{Modeling unit-to-unit variability}
\label{sec:unit-to-unit}

In Sections~\ref{sec:GP}--\ref{sec:noisy_gp}, we discussed how to model a single degradation path using a noisy gamma process. In practice, however, we often measure the degradation of a population of $m$ units that have the same specifications, yet whose degradation paths exhibit heterogeneity. \citet{Meeker_2022} refer to such tests as `repeated-measures degradation tests' (RMDT). One way of modelling such data is to assume that the degradation paths of all units are realisations of the same gamma degradation process, that is, a gamma process with the same $\mu$ and $\nu$. However, this assumption may be too restrictive since there may be additional variability between units that results from slight variations in operating conditions or manufacturing conditions. The most common way to model this heterogeneous degradation beyond what can be explained by the variability of the gamma degradation process and any covariates is to use a `mixed effects' model, in which some of the parameters of the model---so-called `random effects'---vary between units or individuals, whereas others, the `fixed effects', do not\footnote{This definition is just one of five that \citet{Gelman2005} lists.}. Early examples in the degradation literature include \citet{lu_meeker_1993}and \citet{lawless_covariates_2004}, who incorporated random effects into a general path model and gamma process, respectively. A more recent example is \citet{rodriguez-picon_degradation_2018}, who modelled the GaAs laser dataset that was also analysed by \citet{Meeker_1998}. To model the heterogeneity in the degradation paths, \citet{rodriguez-picon_degradation_2018} incorporate random effects into a noise-free gamma process by specifying the effect in either the mean or variance of the gamma process. By contrast, \citet{peng_2018} follow the methodology of \citet{lawless_covariates_2004} and specify random effects in the scale parameter of a gamma process.

In this section, our objective is to demonstrate how the hierarchical model for a noisy gamma process can be extended to incorporate unit-to-unit variability and the advantages of using the mean/coefficient of variation parameterisation. Before we begin, however, it is worth clarifying the terminology that we use. As we pointed out above, the terms random and fixed effects are used when mixed effects models are used to describe unit-to-unit variability. However, as \citet{Gelman2005} and \citet{gelman_hill_2006} point out, all parameters in a Bayesian analysis are random variables; furthermore, because there is a multiplicity of definitions of fixed and random, such terms can engender considerable confusion [Section~6]\citep{Gelman2005}. Consequently, \citet{Gelman2005} and \citet{gelman_hill_2006} make a plea for abandoning these long-used terms in place of more descriptive ones: \emph{varying}, for parameters that differ between groups or units, and \emph{constant}, for parameters that are identical for all groups or units. In this section, we simply identify which parameters are common across units, those that are unique to each unit, and, most importantly, the specification of the prior distribution(s) for parameters that vary from unit-to-unit.

\subsection{Gamma process with unit-to-unit variability}
\label{subsec:gprocess_var}

\begin{figure}[t]
    \centering
    \includegraphics[width=0.8\columnwidth]{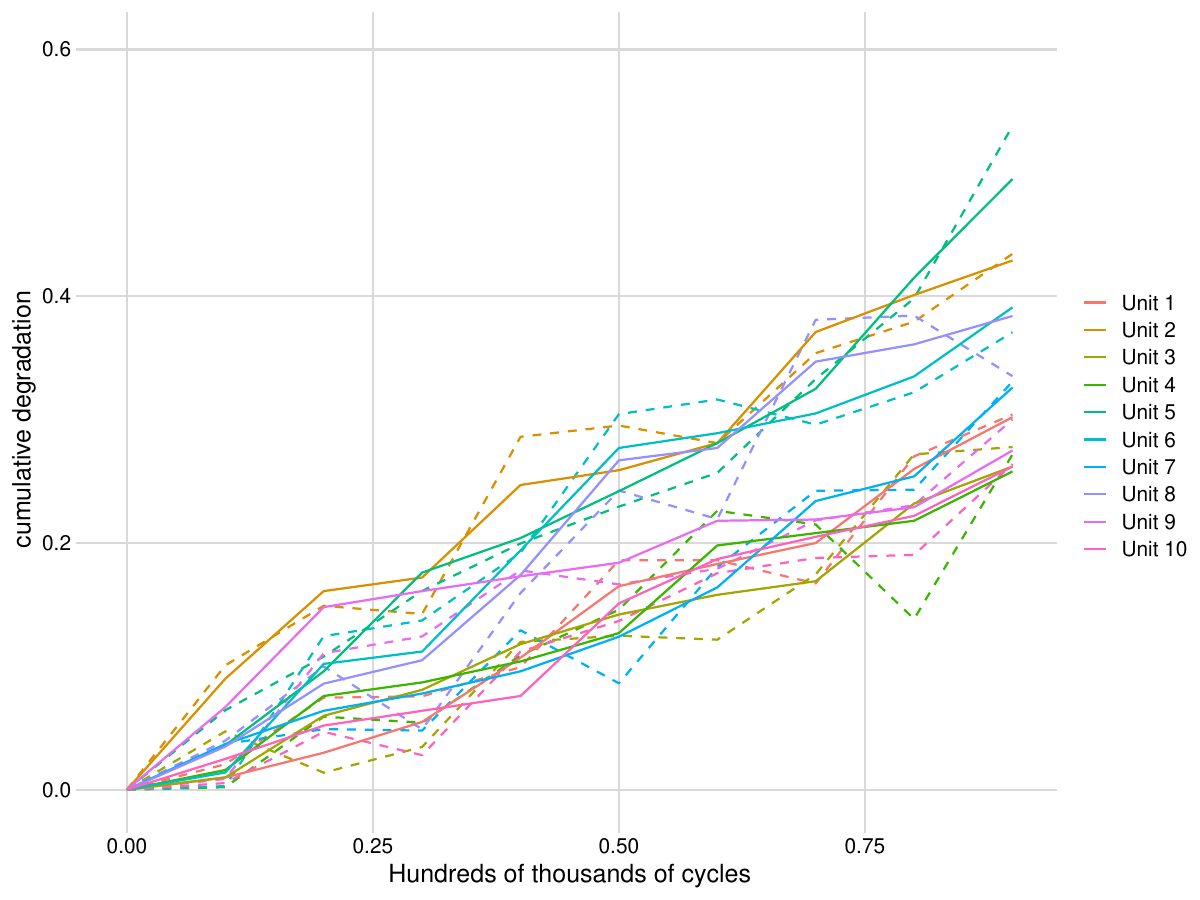}
    \caption{Crack-growth propagation data of \citet{rodriguez-picon_degradation_2018}: original data (solid lines); noisy data (dashed lines).}
    \label{fig:RP_w_noise}
\end{figure}

Figure~\ref{fig:RP_w_noise} shows data from an experiment to measure and then model crack-propagation in the terminal of nominally identical electronic devices \citep{rodriguez-picon_degradation_2018}. The original data are shown as solid lines, to which we have added a small amount of $\mathrm{N}(0, 0.025)$ noise (dashed lines). The soft failure threshold is 0.4~mm, and we can see from the figure that by the end of the experiment, several units have yet to fail. There are two reasons why such data may be collected \citep{robinson_bayesian_2000}: to estimate the remaining useful life or failure time distributions of units that have yet to fail during operation, or the corresponding quantities for new units. In the analysis below, we model the noisy data.

There are three ways in which we might consider modelling the degradation data in Fig.~\ref{fig:RP_w_noise}, each of which leads to a different form of \emph{pooling}, or, alternatively, of how information is shared or not among the units \citep{Johnson_2022}. First, we might make the assumption that the units do not contain information that may be relevant to each other and therefore fit a gamma process separately to each unit; this corresponds to \emph{no pooling}. Second, we might have reason to believe that the observations in Fig.~\ref{fig:RP_w_noise} arise from a single underlying stochastic process, and hence estimate the parameters of this process by averaging the data from all units; this is \emph{complete pooling}. Finally, although the units are different from each other, they have the same specifications, so we might expect their degradation to share similar characteristics: this supposition can be modelled by allowing one or more parameters of the stochastic model to vary from unit-to-unit yet arise from a common distribution. Doing so results in \emph{partial pooling} of information, which is especially useful when sample sizes are small \cite[Section~13.1]{McElreath_2020}. For example, \citet{lawless_covariates_2004} allow the rate parameter of the gamma process to vary from unit-to-unit; in addition, these unit-specific rate parameters arise from a common distribution whose parameters are estimated from the data.

A hierarchical model that pools information from different units provides a principled and coherent framework for estimating the remaining useful life or failure time distributions of units that have yet to fail \textit{and} those corresponding quantities for new units. By contrast, a \textit{no pooling} model cannot, without the use of heuristics, provide information on new units. 

\subsection{Model specification}
\label{subsec:modelspec}

We denote by $y_{ij}$, $j = 1, 2, \ldots, J$, the measured degradation of $J$ identical units, and without loss of generality, we assume that they are measured at the same times $t_i$, $i = 1, 2, \ldots, I$. The three models outlined above can be written using the mean/coefficient of variation parameterisation as described below. In the following exposition, we write some of the priors as joint distributions for notational convenience only, fully recognising that in practice we specify independent prior distributions of the parameters. 

\paragraph{No pooling} In the no pooling model, we assume that separate stochastic processes give rise to the degradation of the $J$ units. Thus, we write the data and process models as 
\begin{align*} 
    y_{ij}|z_{ij}, \sigma_j & \sim \mbox{N}(z_{ij}, \sigma_j)  && \mbox{data model} \\
    \Delta z_{ij}|\mu_j, \nu_j & \sim \mbox{Ga} \left( \frac{\Delta t_{i}}{\nu_j^2}, \frac{1}{\mu_j \nu_j^2} \right) && \mbox{process model} \\
    \mu_j, \nu_j, \sigma_j & \sim \prescript{}{\mu,\nu,\sigma}{\pi}_j (\theta_j) && \mbox{parameter model} 
\end{align*}
where $\prescript{}{\mu,\nu,\sigma}{\pi}_j(\theta_j)$ represents the joint distribution of the parameters of the model for the $j$th unit. The key point to recognise here is that $J$ separate models are fitted to the data and no information is shared among units.

\paragraph{Complete pooling} In the complete pooling model, the degradation paths from $J$ units are assumed to be realizations of the same gamma process. The parameters of this stochastic process are estimated using data from all units, and therefore the BHM can be written as
\begin{align*} 
    y_{ij}|z_{ij}, \sigma & \sim \mbox{N}(z_{ij}, \sigma)  && \mbox{data model} \\
    \Delta z_{ij}|\mu, \nu & \sim \mbox{Ga} \left( \frac{\Delta t_{i}}{\nu^2}, \frac{1}{\mu \nu^2} \right) && \mbox{process model} \\
    \mu, \nu, \sigma & \sim \prescript{}{\mu,\nu,\sigma}{\pi}(\theta) && \mbox{parameter model} 
\end{align*}
In form, the model for complete pooling is identical to the no pooling model, except that because a single gamma process is fitted to all of the data, the unit-specific subscript $j$ is dropped from the parameters $\mu$ and $\nu$.

\paragraph{Partial pooling} Allowing one or both of the parameters of the stochastic process to vary between units, and selecting appropriate hierarchical prior for those parameters, yields a partial pooling model in which we can learn about the degradation of one unit from the degradation of the other units. \citet{lawless_covariates_2004} allow the rate parameter of a gamma process to vary between units but remark that the shape parameter can do so as well; in our reformulation of the gamma distribution, either the mean $\mu$ or coefficient of variation $\nu$, or both, can vary between units. As we pointed out in Section~\ref{subsec:param}, however, $\mu$ and $\nu$ have more intuitive interpretations for practitioners than the shape and rate, and therefore we can more easily decide what information is to be shared. For example, if the units have similar specifications and we therefore believe that we can learn about the mean degradation rate of a unit from the mean degradation of other units, we should allow $\mu$ to vary between units. An equivalent interpretation holds for the coefficient of variation. 

The model below shows only varying $\mu$, but the other partial pooling models follow in a straightforward manner. 
\begin{align*} 
    y_{ij}|z_{ij}, \sigma & \sim \mbox{N}(z_{ij}, \sigma)  && \mbox{data model} \\
    \Delta z_{ij}|\mu_j, \nu & \sim \mbox{Ga} \left( \frac{\Delta t_{ij}}{\nu^2}, \frac{1}{\mu_j \nu^2} \right) && \mbox{process model} \\
    \nu, \sigma & \sim \prescript{}{\nu,\sigma}{\pi}(\theta) && \mbox{parameter model}  \\
    \mu_j & \sim \prescript{}{\mu}{\pi}(\tilde{\theta}) \\
    \tilde{\theta} & \sim \prescript{}{\tilde{\theta}}{\pi}(\psi)
\end{align*}
This partial pooling model allows the average degradation rate of each unit $\mu_j$ to vary, but they arise from a common distribution $\prescript{}{\mu}{\pi}(\tilde{\theta})$ whose parameters $\tilde{\theta}$ are to be learned from the data. As a consequence, information is pooled, or shared, among units. The parameters $\tilde{\theta}$ are themselves assigned a hyperprior $\prescript{}{\tilde{\theta}}{\pi}(\psi)$. In the next subsection, we specify the distributions used as priors and hyperpriors.

\subsection{Fitting a gamma process with unit-to-unit variability}
\label{subsec:fitunitsmean}

To illustrate fitting a noisy gamma process to the data of Fig.~\ref{fig:RP_w_noise} in which the mean degradation rates $\mu_j$ are different from unit-to-unit but arise from a common distribution, we specify the following parameter model:
\begin{align*}
    \sigma & \sim \mbox{Unif}(0, 10) \\
    \nu & \sim t^{+}_3(0, 0.5) \\
    \mu_j | \mu_\mu, \sigma_\mu & \sim \hbox{N}^{+}(\mu_\mu, \sigma_\mu) \\
    \mu_\mu & \sim \mbox{N}^{+}(1, 0.2) \\
    \sigma_\mu & \sim \mbox{Cauchy}^{+}(0, 1)
\end{align*}
The prior distribution for the $\mu_j$ is a Gaussian hierarchical prior whose hyperparameters $\mu_\mu$ and $\sigma_\mu$ are to be estimated. In this way, information is shared across units. The choice of the remaining priors and hyperpriors is based on the prior predictive checking in Section~\ref{subsec:prior} for a single noisy GP. We use the same prior for $\mu_\mu$ as we did for $\mu$, since $\mu_\mu$ now expresses the average degradation per unit time. Furthermore, we use the same priors for $\nu$ and $\sigma$ since, because they are completely pooled, they have the same effect as in the prior predictive checking in Section~\ref{subsec:prior}. Finally, we use a vague prior for $\sigma_\mu$ following the recommendations of \citet[Chapter~17]{gelman_bayesian_2020}. We take the same approach when constructing the hierarchical prior for $\nu$ below. The priors for the hierarchical models could be refined using further prior predictive checking, but we do not do so in order to simplify the presentation.

\begin{figure}[t]
    \centering
    \includegraphics[width=0.8\columnwidth]{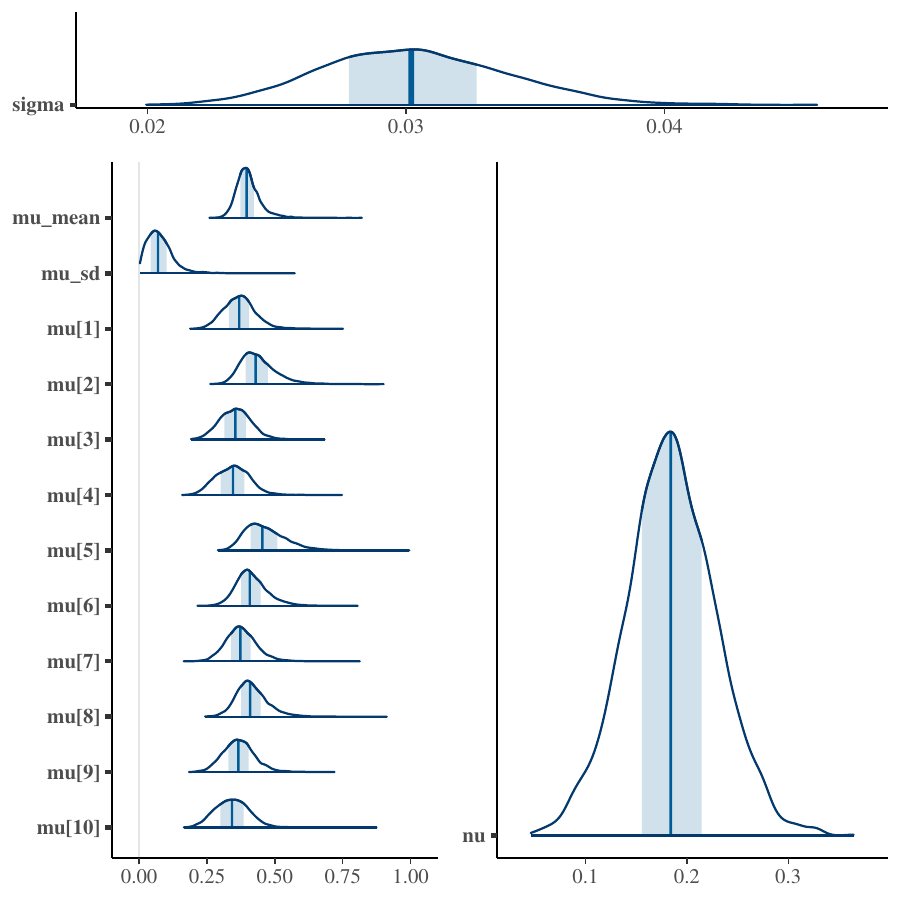}
    \caption{Marginal posterior distributions of parameters from the model where mean degradation rate $\mu_j$ varies from unit-to-unit.}
    \label{fig:pp_mu_marg}
\end{figure}

The Stan code for fitting this model and the resulting output, including diagnostic plots, can be found in Supplementary Material 2. The Hamiltonian Monte Carlo (HMC) algorithm is remarkably efficient, and exploring the posterior distribution requires only 6 chains of length 1000 after a burnin period of 1000 iterations. Table~\ref{tab:pp_mu} shows some summary statistics of the marginal posterior distributions of selected model parameters, and Fig.~\ref{fig:pp_mu_marg} shows the marginal posterior distributions of the parameters $\sigma$, $\nu$, $\mu_1, \ldots, \mu_{10}$, $\mu_{\mu}$ and $\sigma_{\mu}$. In Table~\ref{tab:pp_mu}, the diagnostics $n_{\mbox{eff}}$ and $\hat{R}$ indicate that the chains are well mixed and that estimates of posterior quantiles are reliable \citep{Vehtari_2021}. In addition, few divergences occur. 

\begin{table}
\centering
\caption{\label{tab:pp_mu}Partial output from fitting a BHM to the noisy data of Fig.~\ref{fig:RP_w_noise} where mean degradation $\mu_j$ varies between units. Only statistics for Units~1--4 are shown.}
\centering
\begin{tabular}[t]{lrrrrrr}
\toprule
Parameter & Mean & 2.5\% & 50\% & 97.5\% & $n_{\small{\mbox{eff}}}$ & $\hat{R}$\\
\midrule
\cellcolor{gray!10}{$\sigma$} & \cellcolor{gray!10}{0.03} & \cellcolor{gray!10}{0.02} & \cellcolor{gray!10}{0.03} & \cellcolor{gray!10}{0.04} & \cellcolor{gray!10}{846} & \cellcolor{gray!10}{1.01}\\
$\mu_1$ & 0.37 & 0.26 & 0.37 & 0.49 & 1869 & 1.00\\
\cellcolor{gray!10}{$\mu_2$} & \cellcolor{gray!10}{0.44} & \cellcolor{gray!10}{0.34} & \cellcolor{gray!10}{0.43} & \cellcolor{gray!10}{0.59} & \cellcolor{gray!10}{1646} & \cellcolor{gray!10}{1.00}\\
$\mu_3$ & 0.36 & 0.25 & 0.35 & 0.48 & 1105 & 1.00\\
\cellcolor{gray!10}{$\mu_4$} & \cellcolor{gray!10}{0.35} & \cellcolor{gray!10}{0.23} & \cellcolor{gray!10}{0.35} & \cellcolor{gray!10}{0.47} & \cellcolor{gray!10}{998} & \cellcolor{gray!10}{1.00}\\
\addlinespace
$\nu$ & 0.18 & 0.10 & 0.18 & 0.27 & 263 & 1.02\\
\cellcolor{gray!10}{$\mu_\mu$} & \cellcolor{gray!10}{0.40} & \cellcolor{gray!10}{0.33} & \cellcolor{gray!10}{0.40} & \cellcolor{gray!10}{0.51} & \cellcolor{gray!10}{1766} & \cellcolor{gray!10}{1.00}\\
$\sigma_\mu$ & 0.08 & 0.01 & 0.07 & 0.19 & 456 & 1.00\\
\bottomrule
\end{tabular}
\end{table}

\begin{figure}[t]
    \centering
    \includegraphics[width=0.8\columnwidth]{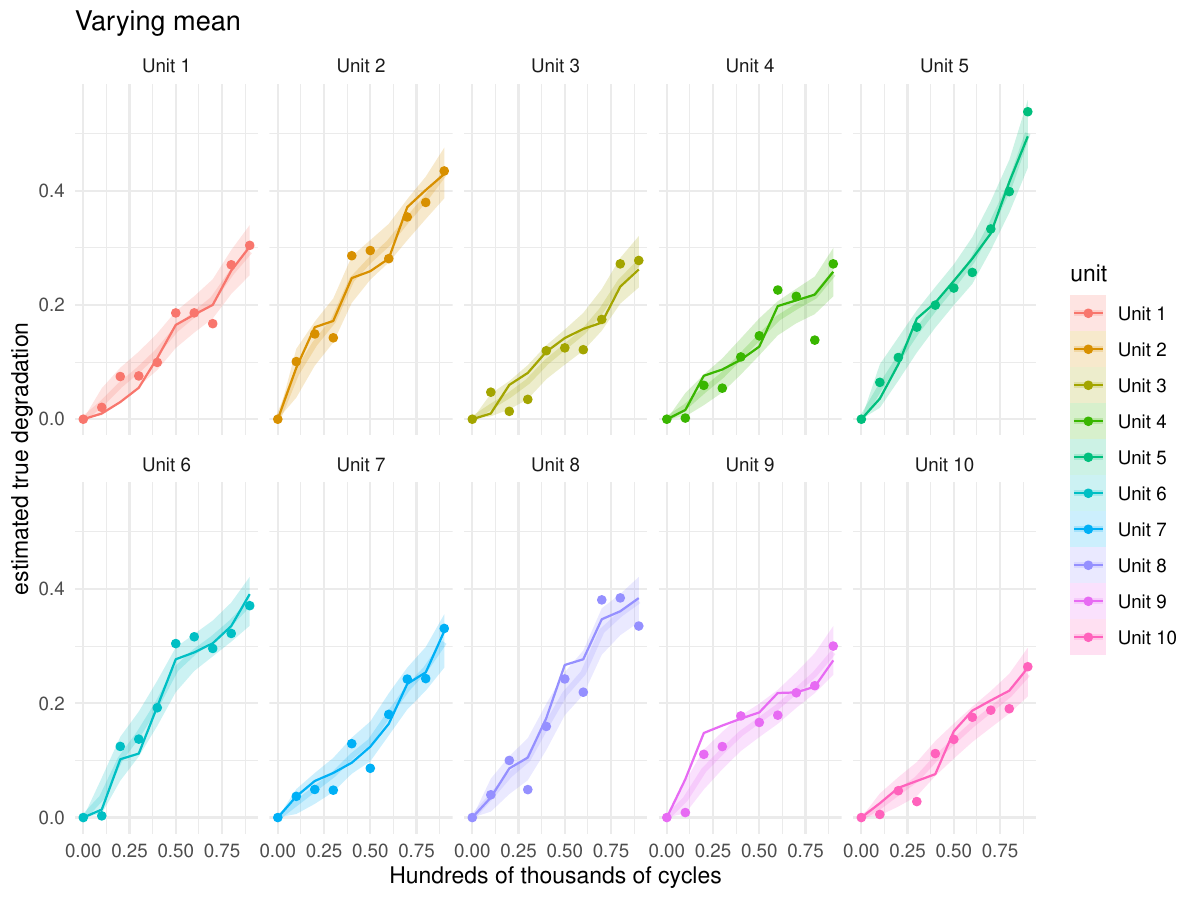}
    \caption{Marginal posterior distributions of the underlying gamma process from a BHM where mean degradation rate $\mu_j$ varies from unit-to-unit. The 0.95 interval of the posterior predictive distribution for each degradation trace is shown as a transparent ribbon, while the true underlying degradation is shown as a solid line and noisy observations are shown as solid points.}
    \label{fig:pp_mu_filtered}
\end{figure}

As Table~\ref{tab:pp_mu} and Fig.~\ref{fig:pp_mu_marg} show, the mean degradation rates of the units arise from the distribution $\mbox{Half-N}(0.40, 0.07)$. The small mean standard deviation of $0.07$ indicates that the mean degradation rates vary in a relatively narrow range, as the posterior distributions in Fig.~\ref{fig:pp_mu_marg} indicate. Despite the noninformative uniform prior of $\sigma$, the measurement uncertainty, the model estimates the mean value to be $0.030$, close to the known value of $0.025$. The BHM also provides posterior distributions of the estimates of the underlying gamma process. These are shown in Fig.~\ref{fig:pp_mu_filtered} as 95\% credible intervals along with the noisy data and the underlying gamma process from which they were generated. As we can see, the credible intervals contain the underlying gamma process over the entire time span for each of the ten units, with few exceptions. 

Using the same BHM structure above, we can also fit a model where the coefficient of variation $\nu$ varies across units and one where both parameters vary. The parameter model of varying $\nu$ model is 
\begin{align*}
    \sigma & \sim \mbox{Unif}(0, 10) \\
    \mu & \sim \mbox{N}^{+}(1, 0.2) \\
    \nu_j | \mu_\nu, \sigma_\nu  & \sim \mbox{N}^{+}(\mu_{\nu}, \sigma_{\nu}) \\
    \mu_\nu & \sim t^{+}_3(0, 0.5) \\
    \sigma_\nu & \sim \mbox{Cauchy}^{+}(0, 0.2)
\end{align*}
and to fit a BHM where both $\mu$ and $\nu$ vary across units, we incorporate the conditional priors for $\mu_j$ and $\nu_j$ and the prior for $\sigma$ into the parameter model.

Table~\ref{tab:pp_nu} shows the results of fitting the varying-$\nu$ model. We can see that the posterior distributions of the $\nu_j$ are almost identical, differing only in their upper-tail behaviour. Complete results for this and the model where both $\mu$ and $\nu$ vary are shown in Supplementary Material 2.

\begin{table}
\centering
\caption{\label{tab:pp_nu}Partial output from fitting a BHM to the noisy data of Fig.~\ref{fig:RP_w_noise} where the coefficient of variation $\nu_j$ varies between units. Only statistics for Units~1--4 are shown.}
\centering
\begin{tabular}[t]{lrrrrrr}
\toprule
Parameter & Mean & 2.5\% & 50\% & 97.5\% & $n_{\small{\mbox{eff}}}$ & $\hat{R}$\\
\midrule
\cellcolor{gray!10}{$\sigma$} & \cellcolor{gray!10}{0.03} & \cellcolor{gray!10}{0.02} & \cellcolor{gray!10}{0.03} & \cellcolor{gray!10}{0.04} & \cellcolor{gray!10}{5116} & \cellcolor{gray!10}{1.00}\\
$\nu_1$ & 0.21 & 0.11 & 0.21 & 0.32 & 2274 & 1.00\\
\cellcolor{gray!10}{$\nu_2$} & \cellcolor{gray!10}{0.22} & \cellcolor{gray!10}{0.14} & \cellcolor{gray!10}{0.22} & \cellcolor{gray!10}{0.33} & \cellcolor{gray!10}{2896} & \cellcolor{gray!10}{1.00}\\
$\nu_3$ & 0.22 & 0.14 & 0.22 & 0.34 & 2555 & 1.00\\
\cellcolor{gray!10}{$\nu_4$} & \cellcolor{gray!10}{0.23} & \cellcolor{gray!10}{0.14} & \cellcolor{gray!10}{0.22} & \cellcolor{gray!10}{0.34} & \cellcolor{gray!10}{2503} & \cellcolor{gray!10}{1.00}\\
\addlinespace
$\mu$ & 0.39 & 0.33 & 0.39 & 0.46 & 5561 & 1.00\\
\cellcolor{gray!10}{$\mu_\nu$} & \cellcolor{gray!10}{0.22} & \cellcolor{gray!10}{0.15} & \cellcolor{gray!10}{0.22} & \cellcolor{gray!10}{0.31} & \cellcolor{gray!10}{2004} & \cellcolor{gray!10}{1.00}\\
$\sigma_\nu$ & 0.03 & 0.00 & 0.03 & 0.10 & 1120 & 1.01\\
\bottomrule
\end{tabular}
\end{table}

As a baseline against which to assess the models with varying $\mu$ and $\nu$, we also fit the complete pooling model, in which the $J$ paths are assumed to arise from the same gamma process; hence, $\mu$ and $\nu$ do not vary. The parameter model is
\begin{align*}
    \sigma & \sim \mbox{Unif}(0, 10) \\
    \mu & \sim \mbox{N}^{+}(1, 0.2) \\
    \nu & \sim t^{+}_3(0, 0.5)
\end{align*}
and the output is summarized in Table~\ref{tab:cp}. Comparing the results in Tables~\ref{tab:pp_mu}--\ref{tab:cp}, we can see that the posterior means of $(\mu, \nu)$ from the complete pooling model are almost identical to the posterior means of $\mu_\mu$ and $\mu_\nu$ from the varying mean and varying coefficient of variation models, respectively.

\begin{table}
\centering
\caption{\label{tab:cp}Output from fitting a model with complete pooling to the noisy data of Fig.~\ref{fig:RP_w_noise}. We assume that the data from all units is a manifestation of a single underlying gamma process, and hence the mean and coefficient of variation of the process do not vary from unit-to-unit.}
\centering
\begin{tabular}[t]{lrrrrrr}
\toprule
Parameter & Mean & 2.5\% & 50\% & 97.5\% & $n_{\small{\mbox{eff}}}$ & $\hat{R}$\\
\midrule
\cellcolor{gray!10}{$\sigma$} & \cellcolor{gray!10}{0.03} & \cellcolor{gray!10}{0.02} & \cellcolor{gray!10}{0.03} & \cellcolor{gray!10}{0.04} & \cellcolor{gray!10}{2506} & \cellcolor{gray!10}{1}\\
$\mu$ & 0.39 & 0.34 & 0.39 & 0.46 & 7085 & 1\\
\cellcolor{gray!10}{$\nu$} & \cellcolor{gray!10}{0.21} & \cellcolor{gray!10}{0.15} & \cellcolor{gray!10}{0.21} & \cellcolor{gray!10}{0.30} & \cellcolor{gray!10}{673} & \cellcolor{gray!10}{1}\\
\bottomrule
\end{tabular}
\end{table}

\subsection{Model comparison}
\label{subsec:modcomp}

Having fitted several models to the crack-growth data in Fig.~\ref{fig:RP_w_noise}, the next step is to check how well they describe the data generating process and to compare them. To do so, we evaluate their ability to predict new observations. In the absence of an independent, external test set, it is conventional to use \textit{information criteria} to compare models. These criteria, such as AIC, DIC, and others, are used to seek a compromise between goodness-of-fit and model complexity and to assess out-of-sample prediction accuracy. AIC and DIC are easy to calculate, but they are not fully Bayesian; hence, criteria such as WAIC (Watanabe-Akaike Information Criterion) and cross-validation ($\hbox{CV}$) are to be preferred \citep{Vehtari2017}.

According to \citet{Vehtari2017}, a useful measure of predictive accuracy for future data points is the expected log pointwise predictive density ($\hbox{elppd}$). It is the expected log likelihood of a new dataset of equivalent size given the posterior conditioned on the observed dataset. It can be written as 
\begin{equation} \label{eq:elppd}
    \hbox{elppd} = \sum^I_{i = 1} \int \log p(\Tilde{y}_i|y) f(\Tilde{y}_i) d\Tilde{y}_i,
\end{equation}
where the summation is over $I$ future observations $\{\Tilde{y}_1, \Tilde{y}_2, \ldots, \Tilde{y}_I\}$ (the same number as in the observed data), $y$ represents the observed data, $\log p(\Tilde{y}_i|y)$ is the likelihood of a new observation $\tilde{y}_i$ given the posterior predictive density of the parameters and intermediate quantities, and $f(\Tilde{y}_i)$ is the distribution of the true generative process for $\Tilde{y}_i$.

In practice, however, we do not know the true data generating process $f(\Tilde{y}_i)$ in eq.~(\ref{eq:elppd}). Nevertheless, an approximation to eq.~(\ref{eq:elppd}) can be calculated for a Bayesian model by using cross-validation whereby we iteratively withhold a portion of the observed data, sample from the posterior conditioned on the remaining data, and then calculate the log likelihood of the withheld portion of the data given the sampled values of the parameters from the posterior. In leave-one-out cross-validation ($\hbox{LOO-CV}$), the size of the withheld portion of the data is equal to one observation, and each of these observations is held out one at a time. The corresponding expression for computing this approximation to $\hbox{elppd}$, which we denote by $\hbox{elppd}_{\hbox{\tiny{LOO-CV}}}$, can be written as
\begin{equation} \label{eq:elppd_loo}
    \hbox{elppd}_{\hbox{\tiny{LOO-CV}}} = \sum^I_{i = 1} \log \frac{1}{S} \sum^S_{s = 1} p(y_i|\theta_{-i}^s).
\end{equation}
In eq.~(\ref{eq:elppd_loo}), $\theta_{-i}^s$ denotes one set of draws from the posterior distribution of the model parameters obtained after fitting the model to all observations \emph{excluding} observation $y_i$. The parameters can also include unobserved latent variables; in the case of the noisy gamma process, the latent variable represents the underlying (filtered) degradation values. The value of $S$, the number of MCMC draws, is chosen to be large enough as to fully cover the posterior distribution. Since the cross-validation approximation requires fitting the model to $I-1$ observations, it will be slightly biased compared to $\hbox{elppd}$ calculated by conditioning on all $I$ observations. This bias can be estimated by using the method described in \citet{gelman_bayesian_2020} and the estimate of $\hbox{elppd}$ subsequently corrected. However, in cases where $I \geq 10$, as it is for the data analysed here, the bias usually small compared to the variance of the estimate \citep{Vehtari2017}, and hence it can be ignored.

When fitting models to nested data, what constitutes a `new observation' is determined by what aspect of the model the analyst wishes to evaluate. In the hierarchical models that we have constructed, we are concerned with two aspects: first, a model's ability to predict the degradation of completely new, unseen units, and second, its ability to predict the future degradation of units currently under test. The ability to predict the degradation of new units can be estimated from eq.~(\ref{eq:elppd_loo}) by treating the $I$ observations from each unit $j$ collectively as a single new observation, i.e., $y_j = (y_{1j}, \ldots, y_{Ij})$. We call this method leave-one-unit-out cross-validation ($\hbox{LOUO-CV}$). By contrast, the $\hbox{elppd}$ for new observations from the units under test can be approximated using eq.~(\ref{eq:elppd_loo}) by sequentially withholding the final observation, $y_{Ij}$, from each of the $J$ units; we refer to this method as step-ahead cross-validation ($\hbox{SA-CV}$). In both cases, we construct the likelihood of the withheld observations in the same way since we have assumed in the data model that the noisy observations $y_{ij}$ from the different units are independent and normally distributed conditional on the true underlying degradation, $\tilde{z}_{i, j}$, and the measurement error, $\sigma$. However, the definition of the posterior predictive distribution of $\tilde{z}$ that should be used depends on both the cross-validation method ($\hbox{LOUO-CV}$ or $\hbox{SA-CV}$) and the model structure (i.e. complete pooling or partial pooling). The details of constructing the posterior predictive distribution of $\tilde{z}$ for $\hbox{LOUO-CV}$ or $\hbox{SA-CV}$ and the results are outlined below, and the code may be found in Supplementary Material 2.

\paragraph{Leave-one-unit-out cross-validation} 

To calculate $\hbox{elppd}_{\hbox{\tiny{LOUO-CV}}}$, we iteratively withhold the data from each unit $j$, condition on the data from the remaining $J-1$ units and then calculate the log-likelihood of the withheld unit's observations, $y_{j} = (y_{1j}, \ldots, y_{Ij})$. This calculation is based on posterior predictive draws of the withheld unit's filtered degradation path, $\tilde{z}_{j} = (\tilde{z}_{1j}, \ldots, \tilde{z}_{Ij})$, and the posterior draws of $\sigma$. Thus, we can write $\hbox{elppd}_{\hbox{\tiny{LOUO-CV}}}$ as
\begin{equation} \label{eq:elppd_louo}
   \hbox{elppd}_{\hbox{\tiny{LOUO-CV}}} = \sum^J_{j = 1}\sum^{I}_{i = 1} \log \frac{1}{S} \sum^S_{s = 1} p(y_{ij} | \left[\Tilde{z}_{ij}, \sigma \right]_{-j}^s).
\end{equation}
To generate posterior predictive draws of the non-noisy degradation path of a new unit, we sample $I-1$ jumps $\Delta\tilde{z}^s_{ij}$ in degradation from $\hbox{Ga}(\left[\tilde{\mu}_j, \tilde{\nu}_j\right]_{-i}^s)$ and then calculate their cumulative sum to generate the degradation path. If $\mu$ is completely pooled, the $\tilde{\mu}^s_j$ are taken from posterior draws $\mu^s$; similarly, if $\nu$ is completely pooled, the $\tilde{\nu}^s_j$ are posteriors draws $\nu^s$. If, however, the mean degradation varies across units, $\tilde{\mu}^s_j$ is sampled from the (hierarchical prior) $\hbox{N}^{+}(\mu^s_\mu, \sigma^s_\mu)$; in the same way, $\tilde{\nu}^s_j$ would be also sampled from $\hbox{N}^{+}(\mu_\nu, \sigma_\nu)$ if the coefficient of variation varied across units. For the models discussed in Sec.~\ref{subsec:fitunitsmean}, the first column of Table~\ref{tab:elppd_loo} shows the value $\hbox{elppd}_{\hbox{\tiny{LOUO-CV}}}$ calculated in this way.

\paragraph{Step-ahead cross-validation}

Step-ahead cross-validation is carried out by iteratively withholding the most recent observation from each of the units under test, and the $\hbox{SA-CV}$ estimate of $\hbox{elppd}$ is calculated as
\begin{equation} \label{eq:elppd_sa_cv}
   \hbox{elppd}_{\hbox{\tiny{SA-CV}}} = \sum^{J}_{j = 1} \log \frac{1}{S} \sum^S_{s = 1} p(y_{Ij} | \left[\tilde{z}_{Ij}, \sigma \right]_{-\left[ Ij \right]}^s).
\end{equation}
To generate the posterior predictive draws in this case, we sample the jump in degradation for unit $j$ from $\hbox{Ga}(\mu^s_j, \nu^s_j)$ and then add this jump to the posterior draws of $\tilde{z}^s_{I-1,j}$. Where either $\mu$ or $\nu$ are completely pooled, $\mu^s_j$ and $\nu^s_j$ are posterior draws $\mu^s$ and $\nu^s$, respectively; otherwise, we use the draws from the posterior distributions of the unit-specific parameters of the gamma process. The $\hbox{elppd}_{\hbox{\tiny{SA-CV}}}$ scores for each of the different models are shown in the right-hand column of Table~\ref{tab:elppd_loo}.

\begin{table}
\centering
\caption{\label{tab:elppd_loo}Leave-one-out cross-validation statistics for the models fitted in Section~\ref{subsec:fitunitsmean}.}
\centering
\begin{tabular}[t]{lrr}
\toprule
  & $\hbox{elppd}_{\hbox{\tiny{LOUO-CV}}}$ & $\hbox{elppd}_{\hbox{\tiny{SA-CV}}}$\\
\midrule
\cellcolor{gray!10}{complete pooling} & \cellcolor{gray!10}{154.7397} & \cellcolor{gray!10}{15.41020}\\
varying $\mu$ & 152.8165 & 14.17090\\
\cellcolor{gray!10}{varying $\nu$} & \cellcolor{gray!10}{153.4844} & \cellcolor{gray!10}{15.09505}\\
varying $\mu$ and $\nu$ & 154.4508 & 15.17758\\
\bottomrule
\end{tabular}
\end{table}

Both of the cross-validation calculations above involve repeatedly re-fitting the model to different subsets of the data, which is computationally inefficient. A much more efficient method is to approximate the $\hbox{elppd}_{\hbox{\tiny{LOO-CV}}}$ \citep{Vehtari2017}; however, in hierarchical modelling cases when the `left out' portions of the data are nested, and as the size of the nested portions increase, the approximations are less likely to work well \citep{Vehtari2017}. Therefore, we choose to use the full cross-validation scheme and incur the computational overhead.

As the $\hbox{elppd}$ results in Table~\ref{tab:elppd_loo} show, the complete pooling model performs the best in both cases by a small margin. Hence, for the data shown in Fig.~\ref{fig:RP_w_noise}, the complete pooling model with measurement error is sufficient to explain the variability in the degradation traces, which consists of measurement error and the inherent volatility of the gamma process. Adding an extra layer of variability by allowing the mean degradation $\mu$ or the coefficient of variation $\nu$ (or both) to vary between units increases the variability of the posterior predictive distribution and results in slightly lower $\hbox{elppd}$ values.  

In this section we have shown how to fit different noisy gamma process models and demonstrated a principled way of choosing the most appropriate model for the data set being analysed. Once the most appropriate model has been chosen, failure time distributions can be constructed for the units under test as well as for new units, as we show in the next section.

\subsection{Failure time distributions}
\label{subsec:ftdist}

As noted in Section~\ref{subsec:gprocess_var}, degradation measurements are collected so that we can estimate the failure time distribution of individual units that are in-service but have not failed and/or of new units that have the same nominal specifications as the experimental units. In many industrial applications, assets are not operated until they can no longer fulfil their intended function; instead, they are operated until such time as the degradation path crosses a `soft' threshold $z_f$. In the case of pipes, for example, $z_f$ might be the minimum wall thickness that is acceptable for safe operation. 

The failure time $T$ can be defined as the first passage time when the true degradation path crosses the failure threshold $z_f$, that is \citep{balakrishnan_2017},
$$
T = \inf\left[ t|Z_t \geq z_f \right].
$$
Note that it depends on the true degradation path, not on the observed one, and hence it does not involve the measurement error \citep{hamada_2008}. In the Bayesian context, we write the failure time distribution as $F_{T|\Theta}(t)$, where $\Theta$ denotes the (hyper) parameters in the model. It can therefore be written as
$$
F_{T|\Theta}(t) = p(T < t | \Theta) = p(Z_t > z_f | \Theta).
$$
One of the advantages of using a fully Bayesian treatment is that we can use the posterior predictive distribution of the underlying gamma process $Z_t$ to calculate the failure time distribution, thereby incorporating uncertainty in the parameters, which will be reflected in credible intervals for $F_{T|\Theta}(t)$. Alternatives to a fully Bayesian treatment include, for example, bootstrapping \citep{peng_2018}.

Although there is no explicit expression for $F_{T|\Theta}(t)$ it is straightforward to obtain the posterior distribution of $F_{T|\Theta}(t)$ by simulation and numerical evaluation of the distribution function of a gamma distribution (e.g., by using the R function \texttt{pgamma}), using a modified version of the procedure outlined by \citet[Sec.~8.2.1]{hamada_2008}. For the complete pooling and varying $\mu$ models, the algorithms are shown in Table~\ref{fig:FT_algs}.

\begin{table}
\centering
\begin{multicols}{2}

\textbf{Complete pooling}
\begin{enumerate}
    \item Draw a sample from the posterior distribution of $(\mu, \nu)$;
    \item Given that $Z_t|\mu, \nu \sim \mbox{Ga}(t/\nu^2, 1/\mu \nu^2)$, calculate $p(Z_t > z_f)$ numerically for a range of values of $t$ to generate one draw of $F_T(t)$;
    \item Repeat Steps~1. and 2. $n_{\hbox{\small{sim}}}$ times.
\end{enumerate}

\columnbreak

\textbf{Varying $\mu$}
\begin{enumerate}
    \item Draw a sample from the posterior distribution of $(\mu_\mu, \sigma_\mu, \nu)$;
    \item Generate $\mu_j$ from $\hbox{N}^{+}(\mu_\mu, \sigma_\mu)$;
    \item Using the $\mu_j$ from Step~2. and the corresponding value of $\nu$ in Step~1., generate a draw of $F_T(t)$ for a range of values of $t$ as in Step~2. for complete pooling;
    \item Repeat Steps 1.--3. $n_{\hbox{\small{sim}}}$ times.
\end{enumerate}

\end{multicols}
\caption{Algorithms for calculating the posterior distribution of the failure time distribution $F_T(t)$ for the complete pooling and varying $\mu$ models.}\label{fig:FT_algs}
\end{table}

\begin{figure}[b]
    \centering
    \includegraphics[width=0.95\columnwidth]{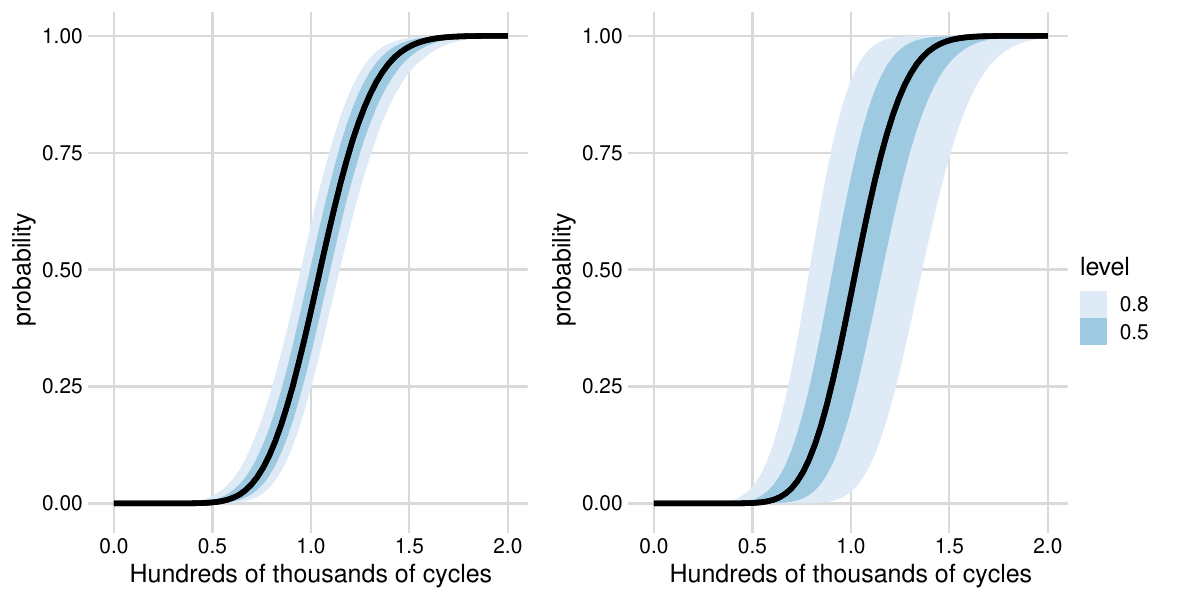}
    \caption{Posterior distributions of the failure time distributions from the complete pooling model (left) and the varying $\mu$ model (right).}
    \label{fig:FT_CP_VM_new}
\end{figure}

Fig.~\ref{fig:FT_CP_VM_new} shows the posterior of the failure time distributions for new units, calculated from the complete pooling and varying $\mu$ models. There is considerably greater uncertainty in $F_T(t)$ from the partial pooling model, but this is not surprising: in addition to the inherent variability of the gamma process, the partial pooling model also includes the variability in the $\mu_j$. 

\begin{figure}[thbp]
    \centering
    \includegraphics[width=0.95\columnwidth]{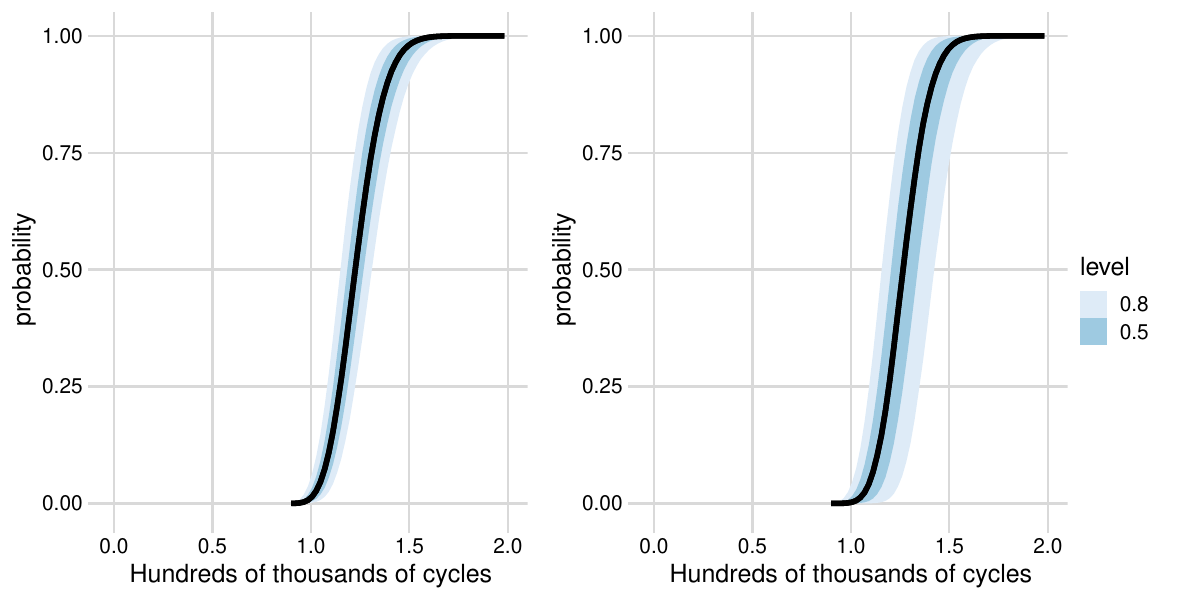}
    \caption{Posterior distributions of the predictive failure time distributions from the complete pooling model (left) and the varying $\mu$ model (right) for Unit~3.}
    \label{fig:FT_CP_VM_U3}
\end{figure}

Using slight modifications of the algorithms shown in Table~\ref{fig:FT_algs}, we can also calculate the failure time distribution for a unit that is currently under test and that has yet to fail, for example, Unit 3. This distribution, also known as the predictive failure time distribution \citep{lawless_covariates_2004}, is conditional on the unit not having failed at some time $t$ and having attained a degradation level $z_t$, which we obtain estimates of during the MCMC sampling routine, $z^s_t$. Figure~\ref{fig:FT_CP_VM_U3} shows the posterior predictive failure time distributions for Unit 3, which has not failed at ; again, we see that because of the additional layer of uncertainty, the credible intervals for the varying $\mu$ model are wider than those from the complete pooling model. Although the unit specific failure time under the varying $\mu$ model is much more precise than the corresponding estimate for a new unit, since we are now using the unit specific estimate $\mu_j$ rather than averaging over the unit-to-unit variability.

\section{Discussion and Conclusions}\label{sec:disc_concl}

The main objective of this paper has been to show that using the Bayesian hierarchical formalism allows us to frame a model for a noisy gamma stochastic process in a tractable and transparent manner. Decomposing the noisy gamma process into a sequence of conditional models---the data, process, and parameter models---removes the need for complex deconvolutions that require the evaluation of, or approximations to, multidimensional integrals. In addition, we showed how the BHM can be naturally extended to include unit-to-unit variability, and how allowing some parameters to vary determines how information is share between observational units. Below, we summarise the main elements of our work and highlight its contributions.

\paragraph*{Reparameterisation of the gamma distribution} Reparameterising the gamma process in terms of the mean $\mu$ and coefficient of variation $\nu$ simplifies specifying prior distributions because they are more easily interpretable than the shape $\beta$ and rate $\xi$: $\mu$ is the mean wear rate, and $\nu$ is the inverse of the `signal-to-noise' ratio, and hence is a measure of the volatility of the gamma process.  Reparameterising the gamma process in this way also helps clarify how unit-to-unit variability can be incorporated into the model---do we expect the mean wear rate to vary among units, or the volatility? In the same way, the new parameterisation also clarifies how we might model the effect of covariates on the degradation: would we expect, for example, environmental variables such as varying temperature or humidity to affect the mean wear or the volatility? Finally, the parameters $\mu$ and $\nu$ are orthogonal, which has desirable computational benefits.

\paragraph*{Identifiability} In Section~\ref{sec:noisy_gp} we demonstrated fitting the noisy GP to simulated data for a single degradation trace. In doing so we identify issues with sampling from the posterior when there are only a few observations. The poor behaviour of the sampling can be attributed to pre-asymptotic non-identifiability between the volatility of the gamma process (expressed by $\nu$) and the measurement error ($\sigma$). Sampling becomes much better behaved when we include more information in the analysis that helps to separate these two parameters, whether that be more data, more informative priors, or modelling the degradation of multiple similar units simultaneously.

We note in passing that in an early paper on noisy Wiener processes, \citet{whitmore_1995} also remarked on the difficulty in estimating the measurement error variance of a noisy Wiener process. In a Bayesian reanalysis of the same data, \citet{hamada_2008} imposed strong prior distributions on the measurement error variance and the variance of the Wiener process in order to ensure identifiability.

\paragraph*{Unit-to-unit variability} For modelling the noisy degradation paths of multiple units simultaneously we explored several models: one where all of the degradation traces arise from the same underlying gamma process (complete pooling) and others where either $\mu$, $\nu$, or both are allowed to vary between units (partial pooling). These models can be seen as nested models, where the varying $\mu$ and $\nu$ model is the most flexible one because both parameters are similar but not identical among units. In models where either $\mu$, $\nu$, or both (as in the complete pooling model) are the same among the different units, this is equivalent to a model in which the hyperparameters $\sigma_\mu$ or $\sigma_\nu \longrightarrow 0$ and the unit specific parameters are forced to be equal to the mean hyper parameters $\mu_\mu$ or $\mu_\nu$.

Based on the $\hbox{elppd}$ criterion, there is a consistent ranking of the models that we fitted, but relatively small differences in the value of the criterion. All of the models fit the data well because the predictive distribution of each unit's underlying degradation path recovers the true underlying degradation path; in addition, the posterior distributions of $\sigma$ includes the true value of $\sigma = 0.025$ that we used to simulate the measurement error. In the posterior distributions of the partial pooling models, there is some evidence that we should allow $\mu$ to vary between units, since there is a some variability in the modes of the marginal posteriors of the unit specific $\mu_j$; however, these distributions are wide enough to encompass the mean $\mu_\mu$ and the posterior of $\sigma_\mu$ has mass near zero and therefore it could well be that they all have the same value, even under a model where we allow them to vary. In the varying $\nu$ models (shown in Supplementary Material 2) the marginal posterior distributions of the unit specific $\nu_j$ are very similar, and $\sigma_\nu$ has considerable mass near zero, showing that there is little evidence that the coefficient of variation varies among the units.

Given the weak evidence for different values of $\mu_j$ in the partial pooling models, it is understandable that the complete pooling model performs best when predicting new data using the two different $\hbox{elppd}_\text{cv}$ methods. Since all models appear to fit the data, the slightly better performance of the complete pooling model is likely because of the reduced uncertainty in the posterior predictive distributions that arises from it being a less flexible model. This is different from the results of \citet{rodriguez-picon_degradation_2018}, who analyse the same data using gamma processes that incorporate unit-to-unit variability but without measurement error; they find that a partial-pooling model slightly outperforms the complete pooling model according to information criteria methods. The difference in outcomes between our analysis here and \citet{rodriguez-picon_degradation_2018} is most likely because the noisy measurements add an extra layer of uncertainty that means that the simpler model slightly outperforms the partial pooling models. Nevertheless, we have shown how analysts can choose between Bayesian models using a fully Bayesian comparison.

\paragraph*{Bayesian workflow and reproducible research} Lastly, we have contributed to the applied Bayesian and reliability literature by demonstrating components of the Bayesian workflow in the reliability domain. These include thinking carefully about how a model can be encoded in order to simplify its implementation; using a contemporary Bayesian computational environment such as Stan; evaluating and refining the model and priors through simulation before fitting the data; interpreting uncertainty in the posterior through posterior predictive distributions; propagating this uncertainty through useful utility functions such as the failure time distribution; and comparing a set of Bayesian models using $\hbox{elppd}$ and cross validation methods. In the interests of reproducible research, we also provide all of the code to implement the models and perform the analysis outlined in this paper in the Supplementary Material and on a Github repository \citep{code_repo} so that others can follow along, reproduce and apply our work in practice.

\newpage

\bibliographystyle{unsrtnat}
\bibliography{references}

\clearpage

\appendix

\section{Diagnostic Plots}\label{sec:ApdxA}

\begin{figure}[h]
    \centering
    \includegraphics[width=0.95\columnwidth]{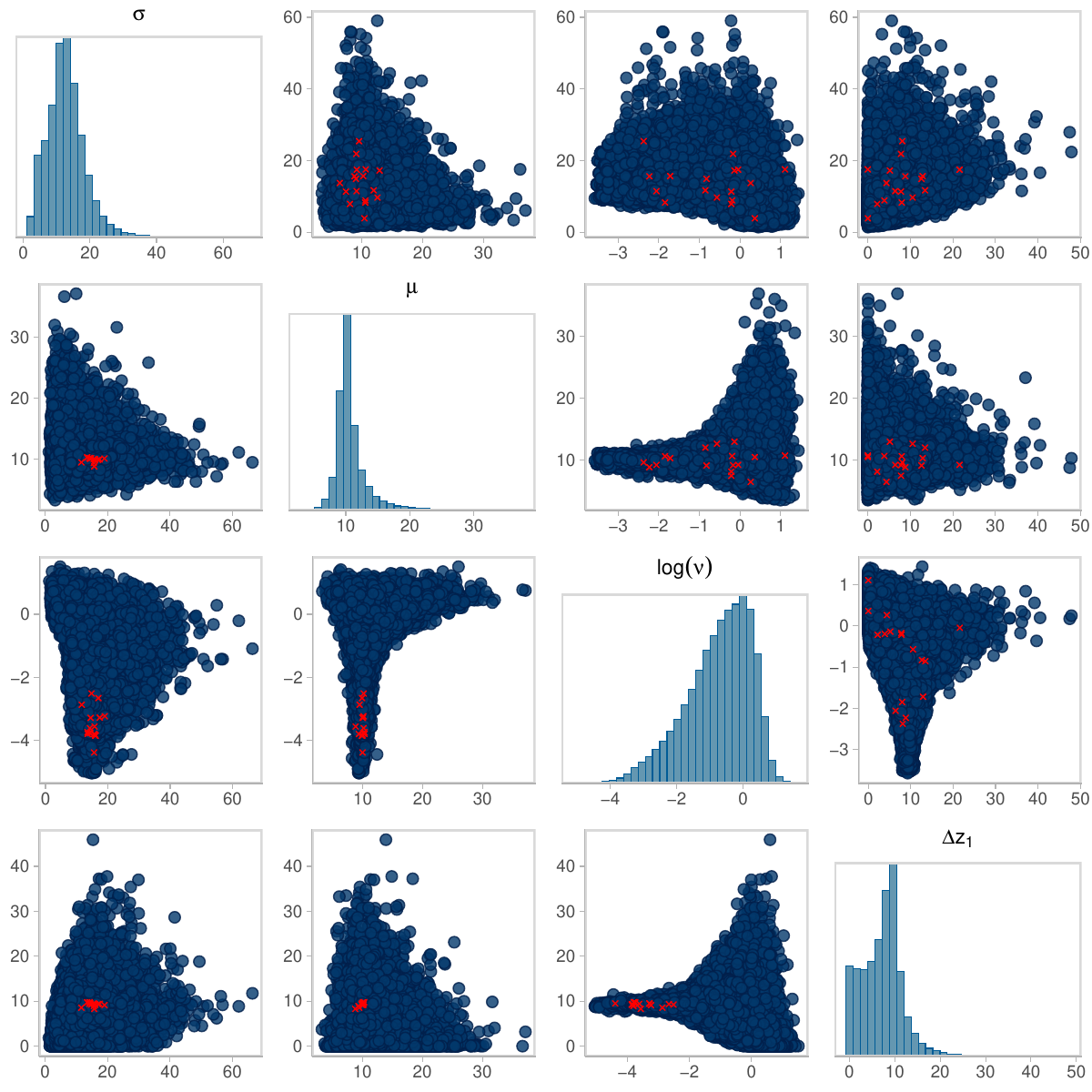}
    \caption{Pairs plot showing the MCMC draws from the posterior distribution of the parameters $\sigma$, $\mu$, $\log\nu$, and the filtered value $\Delta z_1$ when the BHM of the noisy gamma process is fitted to the small dataset of Section~\ref{subsec:datasim}. The red points indicate divergences, which congregate at the end of the funnel in the pairwise plots of $\log\nu$/$\Delta z_1$ and $\log\nu$/$\mu$.}\label{fig:small-data-pairs}
\end{figure}

\end{document}